\documentclass[usenatbib]{mn2e}
\pdfoutput=1

\usepackage{natbib}
\usepackage{amsmath}
\usepackage{graphicx}
\usepackage{rotating}
\usepackage{subfig}
\usepackage{amssymb}
\usepackage{multirow}
\usepackage{times}
\usepackage{commath}
\usepackage{hyperref}
\hypersetup{bookmarksdepth=subsubsection}

\newcommand{\kms}{\textrm{ km s}^{-1}}
\newcommand{\Msol}{\textrm{ M}_\odot}
\newcommand{\dd}{\textrm{d}}
\renewcommand{\vec}[1]{\boldsymbol{#1}}

\begin{document}

\title[Spectroimaging of DG Tau Outflows. I. First epoch]{Multi-epoch Sub-arcsecond [\uppercase{F}\lowercase{e} \uppercase{II}] Spectroimaging of the DG Tau Outflows with NIFS. I. First data epoch}

\author[M.~C.~White et al.]
	{M.~C.~White,$^1$
	 P.~J.~McGregor,$^1$
	 G.~V.~Bicknell,$^1$
	 R.~Salmeron,$^1$
	 T.~L.~Beck$^2$\\
	 $^1$ Research School of Astronomy \& Astrophysics, The Australian National University, Cotter Rd., Weston, ACT, Australia, 2611\\
	 $^2$ Space Telescope Science Institute, 3700 San Martin Dr., Baltimore, MD, USA, 21218}
\date{Accepted 2014 April 2.  Received 2014 April 2; in original form 2013 June 13}
\maketitle

\begin{abstract}

Investigating the outflows emanating from young stellar objects (YSOs) on sub-arcsecond scales provides important clues to the nature of the underlying accretion-ejection process occurring near the central protostar. We have investigated the structures and kinematics of the outflows driven by the YSO DG Tauri, using the Near-infrared Integral Field Spectrograph (NIFS) on Gemini North. The blueshifted outflow shows two distinct components in [Fe II] 1.644 $\mu$m emission, which are separated using multi-component line fitting. Jet parameters are calculated for the high-velocity component. A stationary recollimation shock is observed, in agreement with previous X-ray and FUV observations. The presence of this shock indicates that the innermost streamlines of the high-velocity component are launched at a very small radius, $0.01\textrm{--}0.15\textrm{ AU}$, from the central star. The jet accelerates and expands downstream of the recollimation shock; the `acceleration' is likely a sign of velocity variations in the jet. No evidence of rotation is found, and we compare this non-detection to previous counter-claims. Moving jet knots, likely the result of the jet velocity variations, are observed. One of these knots moves more slowly than previously observed knots, and the knot ejection interval appears to be non-periodic. An intermediate-velocity component surrounds this central jet, and is interpreted as the result of a turbulent mixing layer along the jet boundaries generated by lateral entrainment of material by the high-velocity jet. Lateral entrainment requires the presence of a magnetic field of strength a few mG or less at hundreds of AU above the disc surface, which is argued to be a reasonable proposition. In H$_2$ 1-0 S(1) 2.1218 $\mu$m emission, a wide-angle, intermediate-velocity blueshifted outflow is observed. Both outflows are consistent with being launched by a magnetocentrifugal disc wind, although an X-wind origin for the high-velocity jet cannot be ruled out. The redshifted outflow of DG Tau takes on a bubble-shaped morphology, which will be discussed in a future paper.

\end{abstract}

\begin{keywords}
MHD -- stars: individual: DG Tauri, outflows, variables: T Tauri -- techniques: high angular resolution, imaging spectroscopy
\end{keywords}

\section{Introduction}

It is likely that the outflows driven by accreting young stellar objects (YSOs) play a critical role in solving the angular momentum problem of star formation by removing angular momentum from circumstellar disc material. The nature of this coupled accretion-ejection mechanism remains poorly understood. Magnetic fields are almost certainly integral to this process \citep{MO07}, but the ejection mechanism is still a matter of debate. Outflows could be launched from the stellar surface \citep[e.g.,][]{ST94,MP05}, from points near the truncation radius of the disc, as in the X-wind model \citep{Se94}, or from a range of disc radii via magnetocentrifugal acceleration \citep{BP82,PN83}. Indeed, multiple launch mechanisms may act in concert \citep{L03,FDC06,SLH07}.

Determining the nature of the outflow mechanism is critical in order to understand the underlying accretion process \citep{E09}. Magnetic fields are believed to drive these outflows, and they may also be responsible for inducing disc turbulence via the magnetorotational instability \citep[MRI;][]{BH91,B10}. Both of these processes extract angular momentum from the disc, enabling mass accretion onto the central protostar \citep{MO07}. It is therefore important to determine the physical processes that lead to jet launching, and link these with the properties of the resulting outflow.

Direct observation of the jet launching region is not possible with current optical/near-infrared telescope technology. However, constraints on the jet launching mechanism can be inferred from observations of the outflows close to the central star. For low-mass stars, this takes the form of observing the `microjets' of optically-visible classical T Tauri stars (CTTS). These microjets, which make up the first $\sim 200\textrm{--}300\textrm{ AU}$ ($1\farcs 4$--$2\farcs 1$ at 140 pc) of the outflow, are thought to be largely unaffected by ambient gas, as the jet is expected to clear a channel much wider than the jet via a wide bow shock as it emerges \citep{RCC95}. Most models predict that jet collimation and acceleration occur within $\lesssim 50\textrm{ AU}$ of the star \citep{C07}. Significant effort has been expended over the previous two decades observing these YSO microjets at high angular resolution, first with the space-based Hubble Space Telescope (HST), and later with ground-based adaptive-optics (AO) systems \citep[see, e.g.,][and references therein]{Raye07}.	

One of the most intensely studied T Tauri stars is DG Tauri, which drives the HH 158 and HH 702 outflows \citep{MF83,MRF07}. The accretion and outflow rates determined for this object are amongst the highest of any CTTS \citep{Be02}, with accretion rates approaching $10^{-6}\textrm{ }M_\odot\textrm{ yr}^{-1}$ at some epochs \citep{WG01,WH04}. A multi-velocity structure is observed in the first $\sim$ 300 AU of the approaching outflow, consisting of a well-collimated high-velocity flow near the axis of the system, confined within slower, more spatially extended material. The absolute line-of-sight velocities of the high-velocity component (HVC) are in the range $200$--$400\kms$, with the highest-velocity material positioned closest to the central jet axis and showing bright, shock-excited regions \citep[e.g.,][]{Le97,Be00,Pe03b}. The intermediate-velocity component (IVC) typically shows much broader line widths than the HVC, and is centered around a line-of-sight velocity of $\sim 100\kms$ \citep{Pe03b}.

It is important to understand whether the presence of multiple velocity components in the outflow is the result of multiple launch mechanisms and/or locations, or if it can be described through a single outflow model. For example, \citet{Pe03b} suggested a dual-origin model for the DG Tau outflow, combining a magnetospheric jet with a disc wind. However, it was suggested in the same paper that at least part of the DG Tau IVC could be due to entrainment of this disc wind by the HVC. It would also be possible for a single-component jet to exhibit a double-peaked line profile if, for example, the ionisation of the outflow material varied greatly between inner and outer streamlines, as demonstrated by \citet{Pe04} with analytical models of magnetohydrodynamic disc winds. Therefore, higher-quality data on both velocity components, especially regarding spatial positions, accurate radial velocities, and relative intensities between the components, are required in order to constrain these scenarios.

Improved line velocity determination, coupled with spatial information, will also provide improved constraints on jet rotation. Not only would the unambiguous detection of rotation provide direct evidence that the outflows are extracting angular momentum from the circumstellar disc, but it may also be used to place constraints on the launch radius of the outflow, assuming an MHD disc wind scenario \citep{MO07}. Since the first claims of jet rotation in the DG Tau outflow from HST Space Telescope Imaging Spectrograph (STIS) data \citep{Be02}, many CTTS outflows have been investigated for this signature \citep[e.g.,][]{Ce04}, including a repeat investigation of DG Tau \citep{Ce07}. Radial velocity differences observed across the DG Tau jet have been interpreted as rotation \citep{Be02,Ce07} having the same sense as the rotation inferred for the DG Tau circumstellar disc \citep{Tes02}. The claimed rotation in the IVC is consistent with an MHD disc wind launched from a radius of $\sim$ 3 AU \citep{Be02,Pe04}, whilst the velocity differences across the HVC match a disc wind launched from $\sim 0.2\textrm{--}0.5\textrm{ AU}$ under the assumption that the entire outflow is an MHD disc wind \citep{Ce07}. However, if the IVC results at least partially from entrainment, the line-of-sight velocities could be skewed at any one position by the turbulent motions of shocked gas. In a recent observation of the extreme T Tauri star RW Aurigae, \citet{Ce12} found that the apparent rotation signatures in the outflows from that object change direction over time, and occasionally disappear, indicating that other effects overwhelm any rotation signal present. It is therefore important to understand how the velocities of each component are expected to evolve due to the natural progression of the outflow, and compare this with the observational evidence.

We have obtained three epochs of integral-field spectrograph data of the DG Tau system in the $H$-band over a four-year period (2005--2009). Each epoch provides images of the outflows in [Fe II], in particular the 1.644 $\mu$m line, over an approximately 3\arcsec $\times$ 3\arcsec\ field of view. [Fe II] is one of the strongest forbidden lines present in the near-infrared spectrum,  and is less affected by extinction than optical lines \citep{Pe03b}. In this paper, we present the data from the initial observing epoch (2005), and a small amount of data from the 2006 and 2009 observing epochs. In a future paper, we will introduce the full data from the 2006 and 2009 observing epochs, and discuss the time-evolution of the DG Tau outflows in more detail.

The outflows of DG Tau were most recently investigated in [Fe II] emission by \citet{A-Ae11}, using the SINFONI instrument on the Very Large Telescope. Their data, obtained in 2005 Oct, one month prior to our observations, demonstrate the potential of high-angular resolution spectroimaging for explaining the origin of various outflow components. Here we use our significantly longer ($\sim 20\times$) on-source exposure time, our increased sensitivity to extended structure due to our use of a stellar occulting disc, and our resulting higher signal-to-noise ratio, to rigorously separate the emission from different jet components (Appendix \ref{app:Ftest}), and examine the physical parameters of each one in detail.

This paper is organised as follows.
The observations and data reduction methods are described in \S\ref{sec:obsdata}.
The results of the data reduction are detailed in \S\ref{sec:results}.
We analyse and then remove the stellar spectrum from the data cube, revealing the extended emission structure of the DG Tau outflows.
We use multi-component Gaussian line fitting to separate the blueshifted emission into high- and intermediate-velocity components. We analyse each of these components in detail in \S\ref{sec:discuss}.
The blueshifted high-velocity component denotes the high-velocity jet driven by DG Tau. The knot ejection period of DG Tau cannot be conclusively determined from our data; we suggest that knot ejections in this object are less periodic than previously thought (\S\ref{sec:D-blueknots}).
A stationary recollimation shock is detected at the base of the outflow (\S\ref{sec:D-recoll}), which implies that the innermost streamlines of the jet are initially launched at a high velocity, $\sim 400\textrm{--}700\kms$, from a small launch radius, $\sim 0.01\textrm{--}0.15\textrm{ AU}$ (\S\ref{sec:D-launch}).
Following this rapid deceleration, the jet velocity increases beyond the point where magnetocentrifugal acceleration ceases (\S\ref{sec:D-bluepower}), probably as a result of intrinsic velocity variations (\S\ref{sec:D-blueaccel}).
There is no indication of rotation in the jet (\S\ref{sec:D-bluerot}).
The intermediate-velocity blueshifted component emanates from a turbulent entrainment layer which forms between the jet and either the ambient medium, or the wider-angle molecular wind observed in H$_2$ emission (\S\ref{sec:D-blueH2}).
A magnetic field of strength a few hundreds of $\mu$G to a few mG is expected at these heights above the circumstellar disc, and would facilitate this entrainment (\S\ref{sec:D-entrainreq}). 
We summarise these results in \S\ref{sec:concl}.

\section{Observations and Data Reduction}\label{sec:obsdata}

Initial observations of the DG Tau system in the $H$-band (1.49--1.80 $\mu$m) were obtained using the Near-infrared Integral Field Spectrograph (NIFS) on the Gemini North telescope, Mauna Kea, Hawaii, as part of the NIFS commissioning process on 2005 Nov 12 UT. Data were recorded with the ALTAIR adaptive-optics system in natural guide star mode, using DG Tau itself as the adaptive-optics reference star. NIFS is an image-slicing type integral-field spectrograph that achieves a spectral resolving power $R\sim 5400$ in the $H$-band. The NIFS field has a spatial extent of $3\arcsec\times 3\arcsec$, which is split into 29 slitlets that each pass to the spectrograph. This results in individual spaxels of $0\farcs 103\times 0\farcs 045$, with a two-pixel velocity resolution of $\sim 60\kms$ in the $H$-band \citep{Mce03}. A spatial resolution of 0\farcs 11 was achieved during our observations, based on the observed FWHM of a standard star observed immediately after the DG Tau observations. This corresponds to a distance of $15.4\textrm{ AU}$ at the assumed distance to DG Tau of $140\textrm{ pc}$ \citep{E78}. This distance is intermediate between the radius of Saturn's orbit ($9.5\textrm{ AU}$) and that of Uranus ($19.1\textrm{ AU}$). The instrument was set to a position angle of $\textrm{PA}=316^\circ$, so that the horizontal image axis, corresponding to the coarser spaxel dimension, runs along the known direction of the large-scale HH 158 outflow \citep[$\textrm{PA} = 226^\circ$,][]{MF83}. This places the finer sampling perpendicular to the outflow axis. 

A partially transmissive $0\farcs 2$ diameter occulting disc was used to obscure the central star during the observations, allowing for longer exposures with greater sensitivity to extended structure. Eleven 600 s exposures were taken, with DG Tau being recentered behind the occulting disc every two to five exposures. Two $600\textrm{ s}$ sky frames were also obtained, with an offset of 30\arcsec\ in both RA and Dec. The A0 standard star HIP25736 was observed immediately afterwards, to allow for telluric correction and flux calibration. Flux calibration was based on the 2MASS magnitude ($H=7.795$) for HIP25736, and a shape derived from a blackbody function with a temperature of $7000\textrm{ K}$ that was fit to the 2MASS $J-K$ colour. Flat field, arc, and spatial calibration exposures were obtained on the same night. Standard star observations and flat fields were taken with the occulting disc in place as for observations of DG Tau in order to remove fringing effects generated by the 0.5 mm thick silica occulting disc substrate. These flat-field exposures also allowed for approximate correction of the attenuation of the central star caused by the partially transmissive occulting disc. 

Data reduction was performed using the Gemini NIFS {\sc IRAF} package. An average dark frame was subtracted from each object frame and averaged sky frame. The dark-subtracted average sky frame was then subtracted from the dark-subtracted object frame. A flat-field correction was applied to each slitlet by dividing by a normalised flat-field frame. Bad pixels identified from the flat-field and dark frames were then corrected via 2D linear interpolation.

The individual 2D spectra for each slitlet were transformed to a rectilinear coordinate grid using the arc and spatial calibration frames, and the transformed spectra for each slitlet were stacked in the second spatial direction to form a 3D data cube. All spectra were transformed to a common wavelength scale during this step, so that only spatial registration was required in subsequent data reduction steps. The data cubes derived from each object exposure were then corrected for telluric absorption by division with a normalised 1D spectrum extracted from the observations of the telluric standard star. Hydrogen absorption lines intrinsic to the $H$-band spectrum of the A0 standard star were removed using Gaussian fits to the lines. Flux calibration was achieved using a large-aperture 1D spectrum of the same standard star, which was also corrected for telluric absorption. These final object frames were then spatially registered using the position of DG Tau, and median-combined to produce a final data cube.

The location of the central star in the final data cube was required in order to accurately fix a reference point for the outflow. This location was determined by fitting a two-dimensional Gaussian function to an image produced by collapsing the data cube in the wavelength direction, over wavelength ranges chosen to avoid strong emission lines. The fit was made only over those spaxels within $\sim 0\farcs 25$ of the brightest spaxel in the image, and located the position of the star to within $0\farcs 02$ in the outflow direction ($0\farcs 10$ spaxels), and $0\farcs 01$ in the cross-outflow direction ($0\farcs 04$ spaxels). The FWHM of the continuum image of the DG Tau star is $0\farcs 14$.

\begin{table}
\caption{NIFS Observations of DG Tauri, 2005--2009}\label{tab:obs05}

\begin{tabular}{cccc}
\hline
Date & Epoch & \multicolumn{1}{c}{No.~of on-source} & \multicolumn{1}{c}{Telluric} \\
     &       & \multicolumn{1}{c}{exposures}        & \multicolumn{1}{c}{standard star} \\
\hline
2005 Nov 11 & 2005.87 & 11 & HIP25736 \\
2006 Dec 24 & 2006.98 & 9 & HIP25736 \\
2009 Nov 08 & 2009.88 & 6 & HIP26225\textsuperscript{a} \\
\hline
\end{tabular}\newline
All on-source exposures were 600 s.\\
\textsuperscript{a}2MASS $H$-band magnitude: $7.348$. Blackbody function temperature: $9400\textrm{ K}$.
\end{table}

We include a portion of our multi-epoch data in order to further our arguments regarding the knots in the approaching outflow (\S\ref{sec:D-blueknots}). These data were acquired on 2006 Dec 24 and 2009 Nov 08. The data from each epoch were reduced in the fashion described above, with the main difference being the choice of telluric standard star and the number of on-source 600 s exposures taken. These details are provided in Table \ref{tab:obs05}. We reserve a complete analysis of these multi-epoch data for a future paper. Unless explicitly stated otherwise, all data used within this paper are from the 2005 observing epoch.

Similar NIFS observations of the DG Tau system in the $K$-band (1.99--2.40 $\mu$m), but without the occulting disc, were obtained as a part of the same commissioning process on 2005 Oct 26 UT, and have been presented by \citet{Be08}. We make use of these data in this paper.

\section{Results}\label{sec:results}

\subsection{Stellar Spectrum}\label{sec:results-stellar}

\begin{figure*}
\centering
\includegraphics[width=\textwidth]{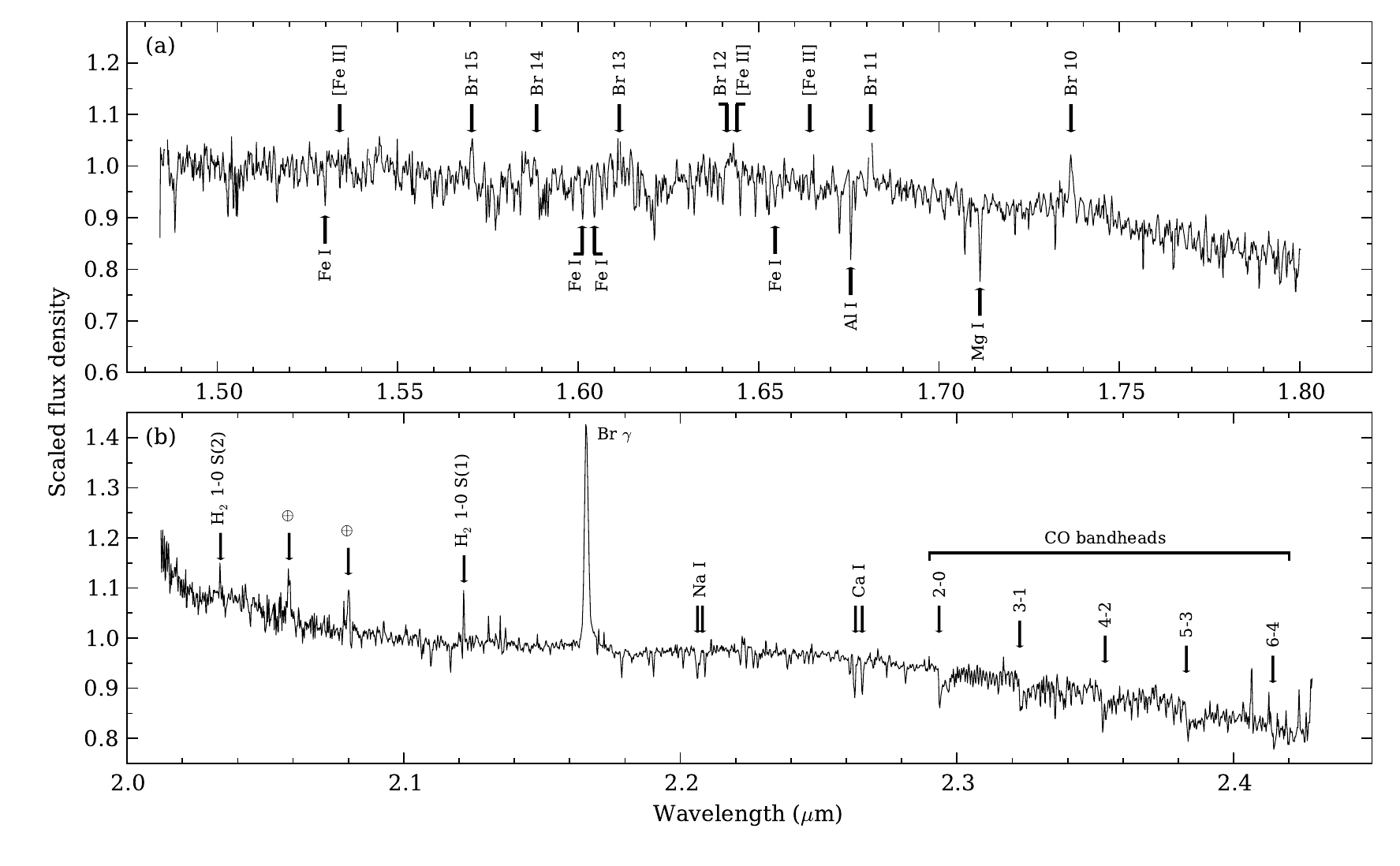}
\caption{The stellar spectrum of DG Tau. The spectrum is extracted from a $0\farcs 8$ diameter circular aperture centered on the star. Panel (a) shows the $H$-band (1.49--1.80 $\mu$m) spectrum, and panel (b) shows the $K$-band (1.99--2.40 $\mu$m) spectrum. The flux density has been normalised to unity at 1.60 $\mu$m in the $H$-band and 2.2 $\mu$m in the $K$-band. Prominent emission features are labelled, as are absorption features, which are used to determine the accuracy of the wavelength calibration (\S\ref{sec:fitted}). The CO bandheads visible in the $K$-band spectrum are also marked. The $K$-band data have been presented previously by \citet{Be08}.}\label{fig:stellar05}
\end{figure*}

Scattered stellar light is apparent across the entire data field. The $H$- and $K$-band stellar spectra of DG Tau were extracted using a $0\farcs 8$ diameter circular aperture, centered on the spatial location of the star in each data cube. It has not been possible to obtain an accurate flux calibration for the $K$-band spectrum as these data were recorded as a flexure test of the NIFS instrument over an extended period in non-photometric conditions. The normalised stellar spectra are presented in Fig.~\ref{fig:stellar05}.

The $H$-band stellar spectrum shows photospheric absorption features, with clearly identifiable K I, Fe I, Al I, and Mg I lines (Fig.~\ref{fig:stellar05}(a)). The $K$-band spectrum shows stellar absorption lines of Na I and Ca I (Fig.~\ref{fig:stellar05}(b)). Previous near-infrared observations of DG Tau on 1994 Dec 14 showed significantly veiled $H$- and $K$-band spectra with few discernible stellar absorption features \citep{GL96}. A similarly veiled spectrum was also seen on 2001 Nov 06 UT \citep{De05}. Furthermore, previous optical observations of DG Tau, where the photospheric spectrum peaks, have shown a highly veiled stellar spectrum, with very few discernible absorption features \citep{HG97}. The source of this veiling continuum is thought to be the accretion shocks occurring close to the stellar surface \citep{Ge00}. Hence, the lack of a veiling continuum indicates DG Tau was in a phase of low accretion activity during the period of our observations.

CO $\Delta v=2$ bandheads are visible in absorption in the $K$-band spectrum (Fig.~\ref{fig:stellar05}(b)), and arise in the stellar photosphere. On the other hand, these bandheads appear in emission in many actively accreting YSOs \citep{C89,C95}. When this occurs, the bandheads typically exhibit a double-peaked structure characteristic of emission from a Keplerian disc, which indicates that the emission arises from the inner radii of the circumstellar disc \citep{C95}. The CO $\Delta v=2$ bandheads in the DG Tau spectrum have been observed to oscillate between appearing in emission \citep{HSR88,C89,Ce93,Be97} and absorption or being absent \citep{GL96, De05}. They also vary significantly in flux, by up to 50\%, on timescales of days \citep{Be97}. The presence of CO bandheads in emission is often associated with an increase in accretion activity, and conversely, the absence, or presence in absorption, of the bandheads is usually associated with a decrease in accretion activity, e.g., the V1647 Orionis outburst of 2003 \citep{RA04,ABR08,Ase09}. Our observation of the DG Tau CO bandheads in absorption provides further evidence that DG Tau was in a low accretion activity phase during the 2005 epoch.

The dominant emission line in the $K$-band spectrum is H I Br $\gamma$ 2.166 $\mu$m. The nature of this line in DG Tau was investigated by \citet{Bee10}. They determined that the majority of the Br $\gamma$ emission emanates from accretion in the circumstellar disc, but approximately $2\%$ of the emission is extended, and coincident with the DG Tau microjet.

\subsection{Stellar Spectrum Removal}

It is necessary to subtract the stellar spectrum and associated spatially unresolved line emission to adequately study the extended emission-line structure of the DG Tau outflows. The $H$-band stellar spectrum shows significant structure in the region of the [Fe II] 1.644 $\mu$m emission line. This consists of a dominant unresolved continuum component, as well as spatially unresolved H I Br 12 emission (Fig.~\ref{fig:stellar05}(a)). $H$-band stellar spectrum subtraction was performed using a custom {\sc python} routine. Our procedure for subtracting the stellar light takes advantage of the orientation of the large-scale DG Tau outflows in the NIFS data cube, and the lack of [Fe II] line emission from the circumstellar disc. Two sample spectra of scattered starlight were formed over a pair of $0\farcs 25$ diameter circular apertures, centered at opposing positions $0\farcs 5$ from the star perpendicular to the outflow direction, and then averaged. For each spaxel, this stellar spectrum was scaled to match the flux observed adjacent to the spectral region of interest for the line being investigated. In the case of the [Fe II] 1.644 $\mu$m line, the region of interest covers a velocity range of  $-380$ to $340\kms$. This scaled stellar spectrum was then subtracted from the spectrum of the spaxel. 

Accurate stellar spectrum subtraction is less important in the $K$-band, due to the less-structured nature of the stellar spectrum in the vicinity of the H$_2$ 1-0 S(1) 2.1218 $\mu$m line. $K$-band stellar spectrum subtraction was performed by forming a pair of continuum images adjacent to the spectral region of interest around the H$_2$ 1-0 S(1) line, averaging them, and subtracting this averaged continuum image from each wavelength plane of the data cube. 

\subsection{Circumstellar Environment}

\begin{sidewaysfigure*}
\centering
\includegraphics[width=0.85\textwidth]{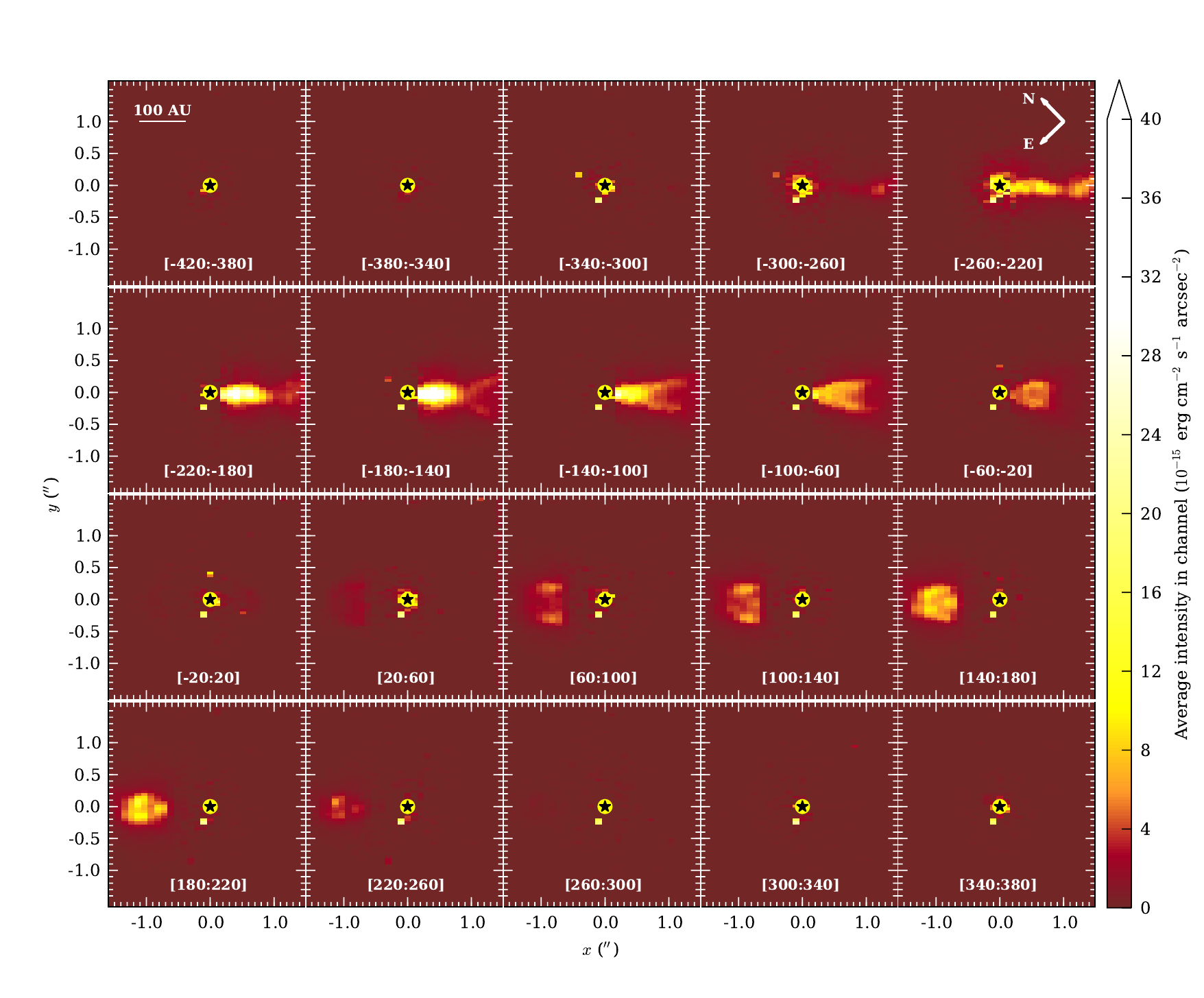}
\caption{Channel maps of the DG Tau outflow. Panels show images of the extended [Fe II] 1.644 $\mu$m line emission around DG Tau, binned into $40\kms$-wide slices. The velocity range of each slice is shown at the bottom of each slice. The velocity ranges used for continuum scaling are also included (top-left and bottom-right panels). The intensity values quoted are the average intensity in each channel over the $40\kms$ velocity range. The black star corresponds to the position of the central star, DG Tau, and the yellow circle indicates the position and size of the $0\farcs 2$ diameter occulting disc.}
\rule{0pt}{0.785\textwidth}
\label{fig:slices05}
\end{sidewaysfigure*}

Fig. \ref{fig:slices05} shows channel maps of the circumstellar environment of DG Tau, as seen in [Fe II] 1.644 $\mu$m line emission, with the stellar and unresolved line emission components removed. The top-left and bottom-right frames show the velocity ranges used for continuum scaling. Here, and in all subsequent figures, the outflow axis is labelled as $x$, and the axis across the outflow as $y$. The data have been binned into $40\kms$-wide slices in order to discern sub-spectral-resolution structure. There are three major outflow components:
\begin{enumerate}
\item A well-collimated, high-velocity blueshifted jet, concentrated in knots of emission. This blueshifted outflow is present in channel maps up to an absolute line-of-sight velocity of $\sim 300\kms$, with the highest-velocity material appearing at the largest observed distance from the central star. This jet has an observed width of $0\farcs 20$--$0\farcs 25\sim 28$--$35\textrm{ AU}$ (approximately the radius of the orbit of Neptune) at the distance of DG Tau; 
\item An intermediate-velocity, less-collimated, edge-brightened, `V'-shaped structure in the blueshifted outflow. Within approximately $1\arcsec$ of the central star, the outer edges of this structure are linear, and are aligned radially with respect to the central star. The opening half-angle of this feature is $15^\circ \pm 1^\circ$. \citet{A-Ae11} obtained an opening half-angle of $14^\circ$ for the same structure from SINFONI data of DG Tau obtained on 2005 Oct 15. This structure is `pinched' $\sim 1\arcsec$ from the star, and then re-expands with increasing distance from DG Tau (Fig.~\ref{fig:slices05}, panel $[-180:-140]\kms$);
\item A redshifted outflow, which  becomes visible approximately $0\farcs 7$ from the central star. The inner region of this structure is obscured by the circumstellar disc around DG Tau. We estimate the radial extent of the obscuration, and hence of the DG Tau circumstellar disc, to be $\sim 160\textrm{ AU}$, after correction for the inclination of the jet-disk system to the line of sight \citep[$38^\circ$;][]{EM98}. This is in agreement with the measurement by \citet{A-Ae11}, who used this obscuration to place limits on the disc models of \citet{ICS10}. The redshifted outflow takes the form of a bubble-like structure. The material with the greatest receding line-of-sight velocity is concentrated on the outflow axis, and at the apex of the bubble. The material along the edges of the structure emits at progressively lower line-of-sight velocities with decreasing distance from the central star.
\end{enumerate}

\subsubsection{Approaching Jet Trajectory}\label{sec:jettraj}

\begin{figure}
\centering
\includegraphics[width=\columnwidth]{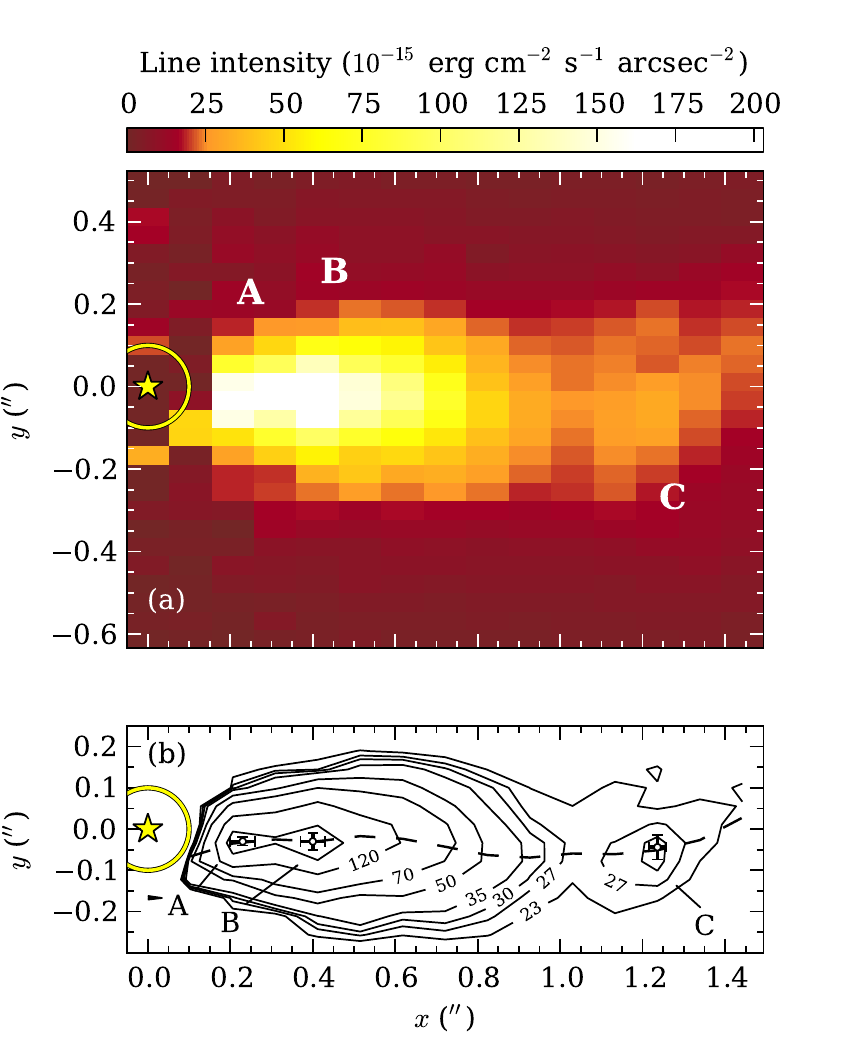}
\caption[DG Tau Approaching Outflow]{The DG Tau approaching outflow. (a) Integrated [Fe II] 1.644 $\mu$m line flux of the approaching outflow from DG Tau. The line flux is computed over the velocity range $-300$ to $0\kms$. Knots A, B and C are labelled. (b) Contour plot of the same integrated [Fe II] line flux. Contours are labelled in units of $10^{-15}\textrm{ erg cm}^{-2}\textrm{ s}^{-1}\textrm{ arcsec}^{-2}$. The unlabelled contour corresponds to $170\times 10^{-15}\textrm{ erg cm}^{-2}\textrm{ s}^{-1}\textrm{ arcsec}^{-2}$. Knots A, B and C are labelled, and the knot centroid positions and associated uncertainties are indicated. The jet ridgeline is shown as a dashed line. The position of the central star and the position and size of the occulting disc used during the observations are shown in both panels by a yellow star and circle, respectively.}\label{fig:blueoutflow}
\end{figure}

The high-velocity blueshifted jet does not travel a linear path, but bends along its length (Fig.~\ref{fig:slices05}, rightmost-top panel). We define the ridgeline of the jet as the location of the jet brightness centre at each position along the outflow axis. Single-component Gaussian fits were performed across the jet, at every recorded position along the outflow axis, on an image of integrated [Fe II] 1.644 $\mu$m emission-line flux formed over the velocity range $-300$ to $-180\kms$. This velocity range was chosen so that the ridgeline was computed for the highest-velocity gas, corresponding to the high-velocity jet \citep[see below, also,][]{Pe03b}. The ridgeline computed from these fits is shown in Fig.~\ref{fig:blueoutflow}(b). The uncertainties in the lateral position of the fitted ridgeline are of order $\pm 0\farcs 01$.

The jet ridgeline was fit in spatial coordinates with a simple sinusoidal function in order to characterise the nature of the jet trajectory. The amplitude of the fitted sinusoid is $0\farcs 027\pm 0\farcs 001\approx 3.8\textrm{ AU}$, and the wavelength is $1\farcs 035\pm 0\farcs 006$. Deprojecting this distance to account for the jet inclination to the line of sight yields a physical wavelength of $235\textrm{ AU}$. If the sinusoidal jet trajectory is due to jet precession, the amplitude corresponds to a precession angle of $\sim 4^\circ$.

\subsubsection{Approaching Jet Knots} \label{sec:knots}

Figs.~\ref{fig:slices05} and \ref{fig:blueoutflow} show that the [Fe II] 1.644 $\mu$m line emission from the blueshifted DG Tau jet is concentrated in a series of three emission knots. We label these features as knots A, B, and C, in order of increasing distance from the central star.\footnote{We choose not to continue the nomenclature of \citet{Pe03b} and \citet{A-Ae11} due to multiple plausible interpretations of the knot ejection history of DG Tau --- see \S\ref{sec:D-blueknots}.} Such emission knots are a common feature of YSO outflows, and have previously been observed in the blueshifted DG Tau outflow on large scales \citep[several arcseconds from the central star;][]{EM98}, as well as on the scale of the microjet \citep[less than $2\arcsec$ from the central star;][]{Ke93, SB93, Le97, De00, Be00, LFCD00, Te02, Pe03, A-Ae11}. With some exceptions \cite[see below, also,][]{LFCD00}, these knots move along the outflow channel at an approximately constant speed.

\begin{table}
\caption{Knot positions in the approaching DG Tau jet --- 2005 epoch}\label{tab:knots05}

\begin{tabular}{ccccc}
\hline
Knot & \multicolumn{1}{c}{Position along} & \multicolumn{1}{c}{Position across} & \multicolumn{1}{c}{Velocity range} & \multicolumn{1}{c}{Centroid [Fe II]} \\
     & \multicolumn{1}{c}{outflow axis}   & \multicolumn{1}{c}{outflow axis}    & \multicolumn{1}{c}{used for fitting} & \multicolumn{1}{c}{line velocity}\\
     & \multicolumn{1}{c}{(\arcsec)}      & \multicolumn{1}{c}{(\arcsec)}       & \multicolumn{1}{c}{($\kms$)} & \multicolumn{1}{c}{($\kms$)} \\
\hline
A & 0.23$\pm$0.03 & -0.03$\pm$0.01 & -260 to -100 & --- \\
B & 0.40$\pm$0.03 & -0.03$\pm$0.02 & -300 to -100 & $\sim 180$ \\
C & 1.24$\pm$0.02 & -0.04$\pm$0.03 & -300 to -180 & $\sim 250$ \\
\hline
\end{tabular}\newline
Quoted uncertainties to the knot positions are the quadrature sum of the fitting errors to the star and knot positions. The fitting uncertainties for knot A are visual estimates. Centroid line velocities are for the high-velocity outflow component (Fig.~\ref{fig:bluevelfit}).
\end{table}

Accurate positions of knots A, B and C relative to the star were determined in order to track their proper motions over time. Two-dimensional spatial Gaussian fits to each knot in integrated [Fe II] 1.644 $\mu$m emission-line flux images were utilised to determine the positions of the knot centroids. The velocity ranges used to form the images for each knot were determined by visual inspection of Fig.~\ref{fig:slices05}. The results of this fitting are presented in Table \ref{tab:knots05} and shown in Fig.~\ref{fig:blueoutflow}. The characteristics of each knot are discussed below.

Knot A is situated $0\farcs 23\pm 0\farcs 03$ along the outflow axis from the central star. The NIFS data Nyquist sample the point spread function across the jet, but undersample the spatial profile in the coarsely-sampled spaxel direction along the outflow. This makes fitting knot A with a two-dimensional Gaussian profile difficult. Visual inspection of these data indicate that the FWHM of the knot is $\sim 0\farcs 1$ in both axes, making it significantly more compact than knots B and C. The difficulty in accurately fitting a Gaussian profile also results in a larger uncertainty in the knot centroid position.

Similar knots at the location of knot A have been observed previously in [S II] \mbox{6716 \AA /}\mbox{6731 \AA}\ \citep{SB93}, [O I] 6300 \AA\ \citep{SB93, Le97} and He I 10830 \AA\ \citep{Te02}. Furthermore, \citet{Le97} report that the emission feature they observe at $\sim 0\farcs 15\approx 34\textrm{ AU}$\footnote{\citet{Le97} report the knot position as $0\farcs 17\pm 0\farcs 05$ from the star in the raw image, and $0\farcs 13$ after deconvolution.} deprojected distance from the central star exhibits very little proper motion, suggesting that the knot represents a steady region in the flow where emission is enhanced. A feature similar to knot A appears to be present in the data of \citet[][fig.~3 therein]{A-Ae11}; however, those authors did not mention it. We interpret knot A as a stationary shock in the jet, resulting from the recollimation of the flow. We expand further on this interpretation in \S\ref{sec:D-recoll}.

Knot B is well-described by a Gaussian profile, which is extended in the outflow direction with an axial ratio of $\sim 2.3$. This knot was most recently detected by \citet{A-Ae11}, who reported a position of $0\farcs 37\pm 0\farcs 03$ along the outflow axis from the central star on 2005 Oct 15. Our positions agree to $1\sigma$. Knot C is significantly fainter than knots A and B, at $\sim 15\%$ of their peak intensity (Figures \ref{fig:slices05} and \ref{fig:blueoutflow}). As with knot B, knot C is elongated in the outflow direction, but with an axial ratio of $\sim 1.7$. This knot was also detected by \citet{A-Ae11}, with a reported position of $1\farcs 2\pm 0\farcs 05$ along the outflow axis from the central star on 2005 Oct 15. This agrees with our measurement to $2\sigma$, although our fitted knot position is within their uncertainties. We conduct an analysis of the recent knot ejection history of DG Tau in \S\ref{sec:D-blueknots}.

\subsubsection{Receding Outflow Morphology}\label{sec:redflow}

\begin{figure}
\centering
\includegraphics[width=\columnwidth]{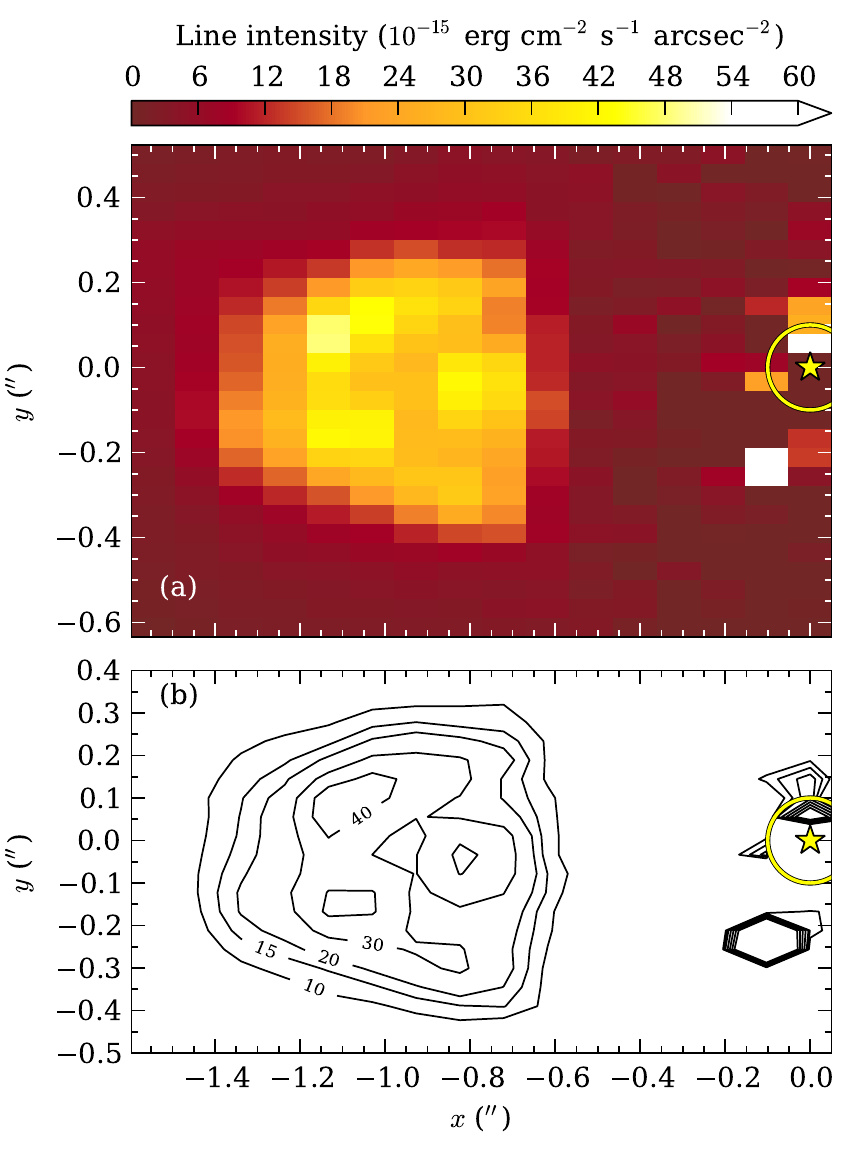}
\caption[DG Tau Receding Outflow]{DG Tau receding outflow. (a) Integrated [Fe II] 1.644 $\mu$m line flux of the receding outflow from DG Tau. The line flux is computed over the velocity range $0$ to $300\kms$. (b) Contour plot of the same integrated [Fe II] line flux. Contours are labelled in units of $10^{-15}\textrm{ erg cm}^{-2}\textrm{ s}^{-1}\textrm{ arcsec}^{-2}$. The position of the central star and the position and size of the occulting disc used during the observations are shown in both panels by a yellow star and circle, respectively.}\label{fig:redoutflow}
\end{figure}

The morphological appearance of the DG Tau redshifted outflow, shown in Fig.~\ref{fig:redoutflow}, is different from that of the blueshifted outflow (Fig.~\ref{fig:blueoutflow}). First, there is no clearly discernible fast outflow, nor ridgeline. Second, the emission from this outflow comes predominantly from a bubble-like structure (Figures \ref{fig:slices05} and \ref{fig:redoutflow}). This structure was observed by \citet{A-Ae11}, and was interpreted as being the redshifted equivalent of a faint `bubble' they claimed in the approaching outflow at similar distances from the central star. We do not observe such a structure in the blueshifted outflow (\S\ref{sec:fitted}), and we will discuss and model the cause of this bipolar outflow asymmetry in a forthcoming paper \citep{MCW13b}.

\subsection{Fitted Line Components}\label{sec:fitted}

\begin{figure*}
\centering
\includegraphics[width=\textwidth]{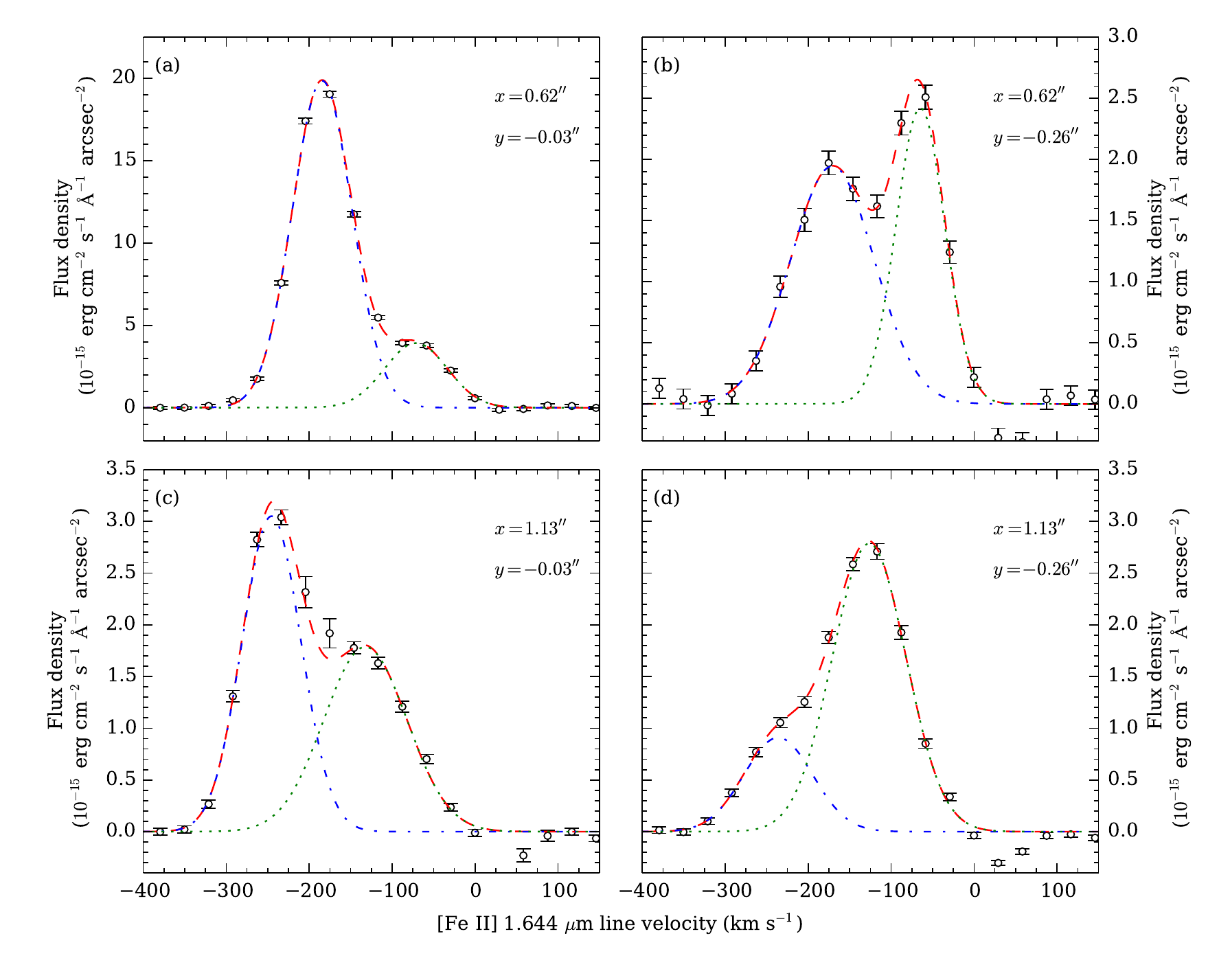}
\caption{Two-component Gaussian fits to the spectra of four spaxels in the approaching DG Tau outflow. The spaxels are located at a distance of (a, b) $x=0\farcs 62$ and (c, d) $1\farcs 13$ along the outflow axis. The spaxels shown in panels (a, c) are on the jet ridgeline at $y=-0\farcs 03$, and the spaxels in panels (b, d) are offset from the ridgeline at $y=-0\farcs 26$. Actual data and uncertainties are indicated by black circles and error bars, fitted line components are shown as blue dot-dashed and green dotted lines, and the total fit is shown as red dashed lines.}\label{fig:spaxspeceg}
\end{figure*}

Visual inspection of our spectra clearly indicates the presence of at least two [Fe II] 1.644 $\mu$m line components at every spatial position with significant signal-to-noise ratio. In many spatial locations, these two components are significantly blended. A multi-component Gaussian fit was performed to separate these spectral components. Both one- and two-component fits were made, and an $F$-test (Appendix \ref{app:Ftest}) was utilised to determine the statistically appropriate number of components to retain in the final fit \mbox{\citep{Wee07}}. Stricty speaking, the use of a likelihood ratio test such as the $F$-test in this situation is statistically incorrect \citep[see Appendix \ref{app:Ftest-applic};][]{Pre02}. However, given the absence of a statistically correct alternative that could be sensibly applied to the number of spectra presented here, and the obvious presence of two line components at most positions, we opt to continue with this approach \citep[e.g.][also see Appendix \ref{app:Ftest-applic}]{Wee12}. Spaxels were excluded from fitting if the signal-to-noise ratio of the brightest spectral pixel in the vicinity of the emission line was less than 10, or if the relative error on the fitted line amplitude and/or width exceeded unity. Spaxels that could not be fit with two components were also excluded. Applying these criteria, it was found that acceptable fits were produced over a region comparable to the detected emission in Figures \ref{fig:slices05}, \ref{fig:blueoutflow} and \ref{fig:redoutflow}. Example spectra, and the fits obtained to those spectra using the above procedure, are shown in Fig.~\ref{fig:spaxspeceg}.

To determine line velocities relative to the systemic velocity, it was necessary to determine the velocity of the central star in our data. To accomplish this, Gaussian profiles were fit to several stellar absorption features in the $H$-band stellar spectrum (Fig.~\ref{fig:stellar05}(a)). The velocity correction obtained was then applied to all line velocities. 

The velocity resolution of our $H$-band data was measured to be $55\kms$, based on Gaussian line fits to observed sky lines. The intrinsic line widths of each fitted profile were determined by quadrature subtraction of this instrumental velocity resolution from the fitted line width, via the formula $\textrm{FWHM}_\textrm{intrinsic}^2=\textrm{FWHM}_\textrm{fitted}^2-\textrm{FWHM}_\textrm{instrumental}^2$. We discuss the properties of each fitted component below.

\subsubsection{Approaching High-Velocity Component}\label{sec:blueoutflowHVC}

\begin{figure*}
\centering
\includegraphics[width=\textwidth]{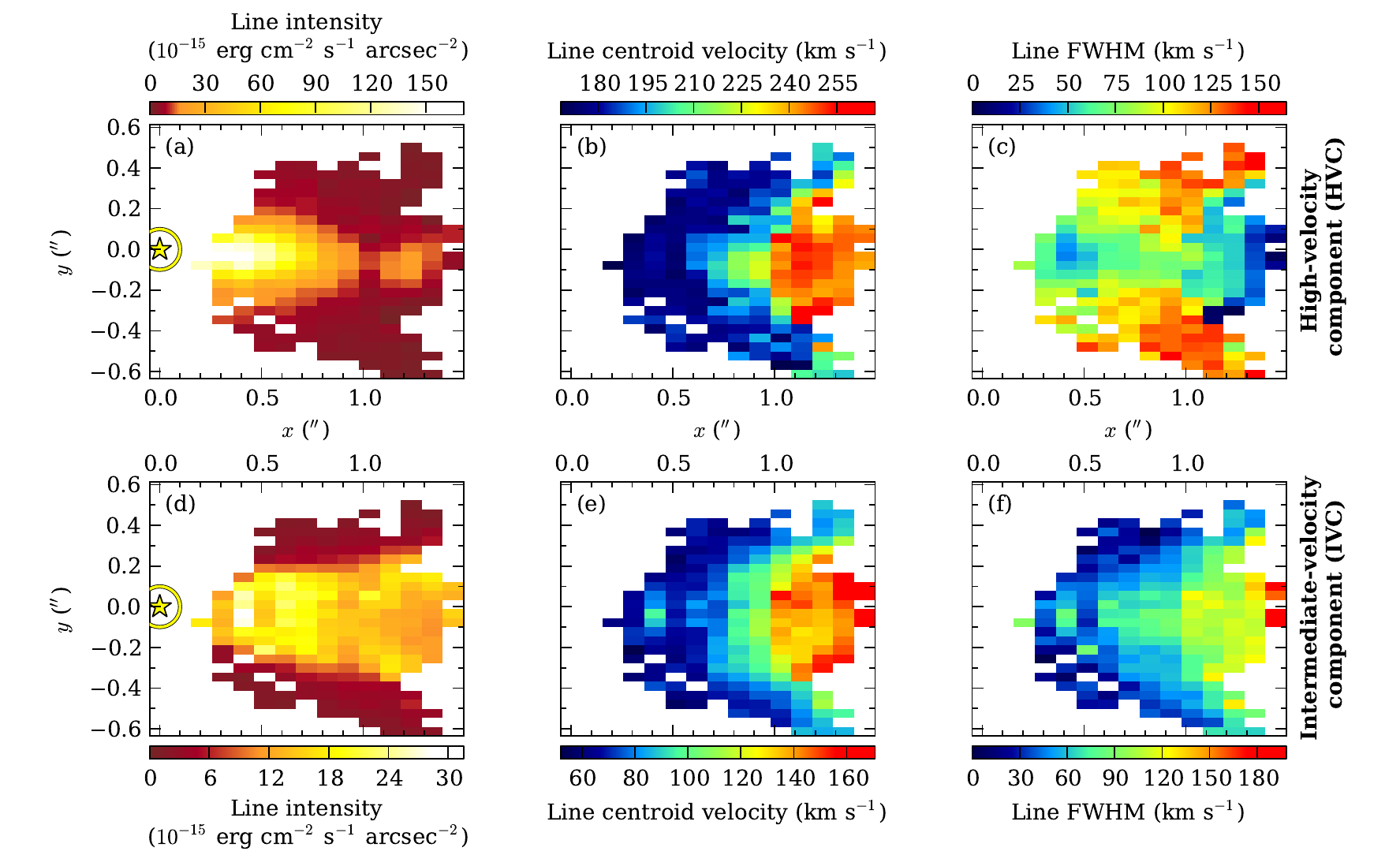}
\caption{[Fe II] 1.644 $\mu$m emission characteristics from the high- and intermediate-velocity components fit to the DG Tau approaching outflow. Panels (a) to (c) show fitted parameters for the high-velocity component. Panels (d) to (f) show fitted parameters for the intermediate-velocity component. Panels (a) and (d) show the fitted line intensity. Panels (b) and (e) show the absolute fitted line velocities, which have been corrected to the systemic velocity of the central star. Panels (c) and (f) show the line FWHM, which has been approximately deconvolved from the instrumental velocity resolution through quadrature subtraction. The yellow star and circle in (a) and (d) represent the position of the star, and the position and size of the occulting disc, respectively.}\label{fig:bluevelfit}
\end{figure*}

The [Fe II] 1.644 $\mu$m emission-line intensity image of the blueshifted HVC in Fig.~\ref{fig:bluevelfit}(a) shows the classic morphology of a well-collimated, high-velocity jet. Knots B and C are reproduced. Knot A is not visible, as that region of the outflow is not fit due to its low signal-to-noise ratio, which results from the proximity of the central star. The `pinching' of the outflow $\sim 1\arcsec$ along the outflow axis from the central star is also reproduced. We interpret this to be due to a lack of emitting gas between the two jet knots, rather than an actual narrowing of the jet.

The peak line velocity at each position along the outflow occurs on the jet ridgeline (Fig.~\ref{fig:bluevelfit}(b)). The line-of-sight line velocity is constant at $\sim 170\kms$ in the region of knot B, $0\farcs 40\sim 91\textrm{ AU}$ deprojected distance from the central star. The peak absolute line velocity increases with distance from the central star between knots B and C. This is in agreement with previous observations of the DG Tau microjet that generally show increasing absolute line velocity with distance from the central star \citep{Be00,Pe03b}, although \citet{Pe03b} shows some evidence for sinusoidal velocity variations (\S\ref{sec:D-blueknots}). The fitted line width is lowest along the jet ridgeline (Fig.~\ref{fig:bluevelfit}(a)), and the region of lowest line width corresponds to the region of highest line component intensity at each position along the outflow axis. This indicates the presence of a narrow jet with a relatively undisturbed core.

\subsubsection{Approaching Intermediate-Velocity Component}\label{sec:blueoutflowIVC}

The integrated line intensity image of the IVC shown in Fig.~\ref{fig:bluevelfit}(d) differs significantly from that of the HVC (Fig.~\ref{fig:bluevelfit}(a)). The emission is spread further from the outflow axis than the HVC. Interestingly, the edge-brightened `V'-shaped structure is not reproduced. This is because the channel maps (Fig.~\ref{fig:slices05}) show intensity over a narrow range of velocities, whilst Fig.~\ref{fig:bluevelfit}(d) displays the total intensity. This indicates that it is the velocity structure that is stratified (Fig.~\ref{fig:bluevelfit}(e)). None of the observed emission knots is reproduced in the IVC. There is a small increase in the IVC line intensity and absolute line velocity at the position of knot B. The small spatial extent of this increase (a few spaxels), and the dominance of the HVC at this position, leads us to conclude that these increases are fitting artefacts.

The IVC velocity structure (Fig.~\ref{fig:bluevelfit}(e)) is similar to the HVC velocity structure but at lower absolute velocities. The IVC line width profile (Fig.~\ref{fig:bluevelfit}) shows a different structure to the HVC, with the regions of highest line width being found on the outflow axis, and the fitted line width decreasing with lateral distance from the outflow axis.

\subsubsection{Receding Outflow}\label{sec:redoutflow}

\begin{figure}
\centering
\includegraphics[width=\columnwidth]{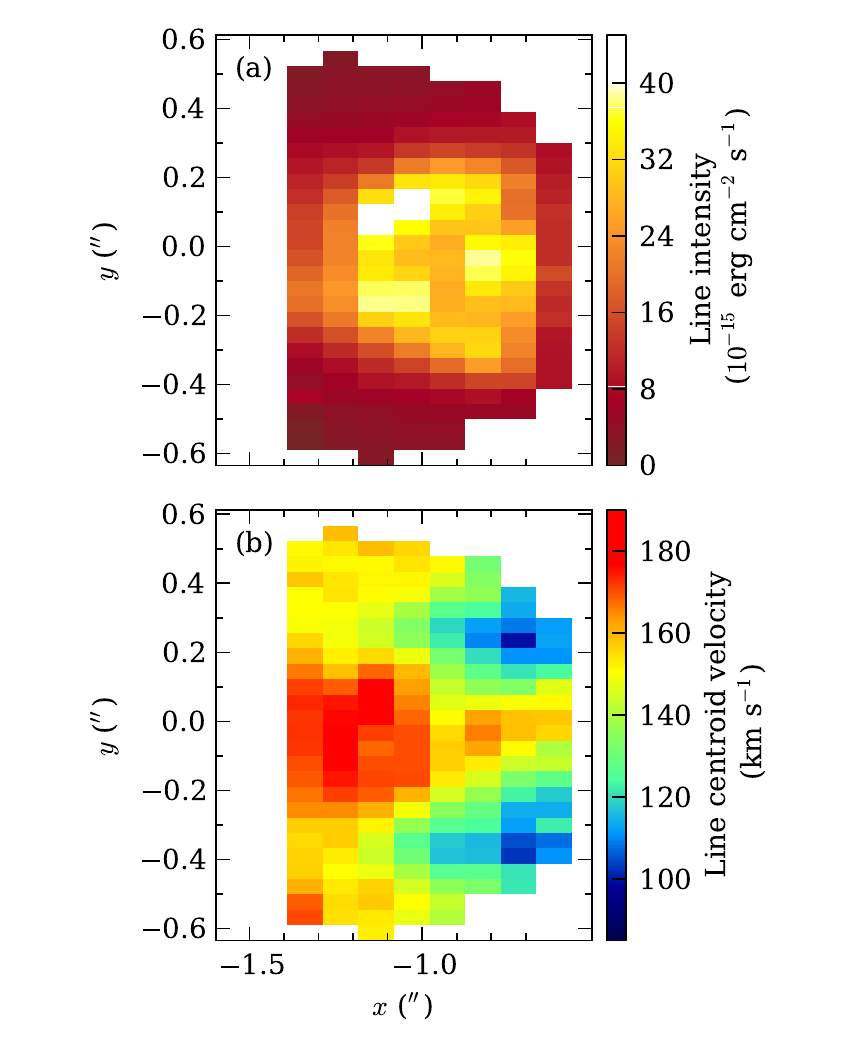}
\caption{[Fe II] 1.644 $\mu$m emission characteristics from the single component fit to the DG Tau receding outflow. Panel (a) shows the fitted line intensity and panel (b) displays the fitted line velocity of the redshifted outflow based on a single-component Gaussian fit. The fitted line velocity has been corrected for the systemic velocity of the central star.}\label{fig:redvelfit}
\end{figure}

The redshifted [Fe II] 1.644 $\mu$m line emission from the receding outflow was fit using the procedure described above. Fits were restricted to a single component, as this is all that is warranted by the data. The resulting fitted [Fe II] 1.644 $\mu$m line component for the receding DG Tau outflow is shown in Fig.~\ref{fig:redvelfit}.

There are several distinctive features in the receding outflow velocity profile shown in Fig.~\ref{fig:redvelfit}(b). The highest line velocities of $\sim 180\kms$ are found at the `apex' of the bubble-like structure, $1\farcs 3$ from the central star. The line velocities of the emission from the structure decrease with decreasing distance from the central star, reaching $\sim 100\kms$ at the edge of the observable emission closest to the star. A ridge of emission with velocity $\sim 160\kms$ runs along the outflow axis for the length of the structure. This suggests that there is an underlying stream of material driving the evolution of this structure. As noted above, we will discuss this further in a future paper \citep{MCW13b}.

\pagebreak

\subsection{Approaching Outflow Electron Density}\label{sec:densne}

The near-infrared lines of [Fe II] arise from low-lying energy levels, and are useful tracers of electron density, $n_\textrm{e}$. In particular, the intensity ratio between the [Fe II] lines at $1.533\textrm{ }\mu\textrm{m}$ and $1.644\textrm{ }\mu\textrm{m}$ provides a diagnostic of electron density in the range $n_\textrm{e}\sim10^2$--$10^6$ cm$^{-3}$ \citep{PZ93}. The derived electron density is only weakly dependent upon the electron temperature, $T_\textrm{e}$, in the range $T_\textrm{e}\sim(0.3\textrm{--}2.0)\times 10^4\textrm{ K}$. We assume an electron temperature of $T_\textrm{e}=10^4\textrm{ K}$ for the DG Tau outflow \citep{B02}. \citet{Pes03} have computed the relation between this line ratio and electron density for a 16-level Fe$^\textrm{+}$ model.

\begin{figure}
\centering
\includegraphics[width=\columnwidth]{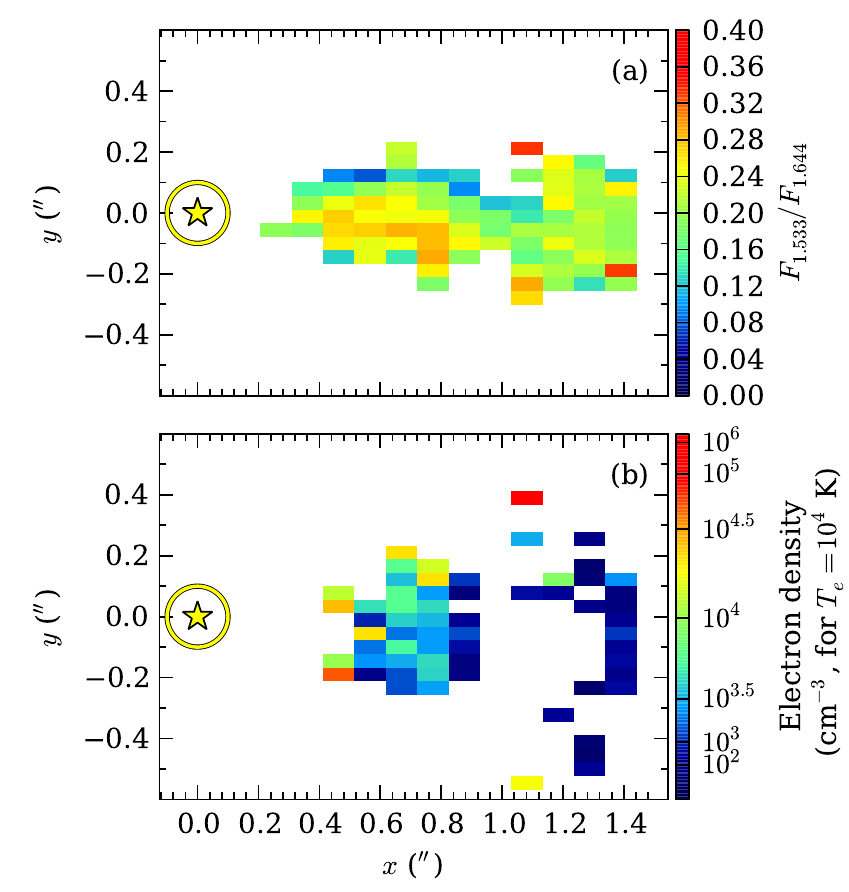}
\caption{Ratio of integrated flux between [Fe II] $1.533\textrm{ }\mu\textrm{m}$ and [Fe II] $1.644\textrm{ }\mu\textrm{m}$ line emission for approaching outflow components. Shown is the computed line ratio for (a) the high-velocity component and (b) the intermediate-velocity component. Line fluxes were determined by integration of the raw, stellar-subtracted spaxel spectra over the velocity ranges $-380$ to $0\kms$, and then splitting the integrated fluxes about the velocity where the two fitted [Fe II] 1.644 $\mu$m line components have equal flux density. Spaxels have been masked where either a threshold signal-to-noise ratio of 5 in the high-velocity component, or two in the intermediate-velocity component has not been reached, or where the ratio value is in the saturation limit for determining electron density \citep[$F_{1.533}/F_{1.644}\gtrsim 0.40$;][]{Pes03}. Electron density has been calculated for an electron temperature of $T_\textrm{e}=10^{4}\textrm{ K}$ over the range of ratios $0.035<F_{1.533}/F_{1.644}<0.395$ (second colourbar). The yellow star and circle represent the position of DG Tau, and the position and size of the occulting disc, respectively.}\label{fig:densrat}
\end{figure}

The [Fe II] $1.533\textrm{ }\mu\textrm{m}/1.644\textrm{ }\mu\textrm{m}$ flux-ratio map of the approaching outflow components derived from our data is shown in Fig.~\ref{fig:densrat}. Integrated line fluxes were determined via integration of the stellar-subtracted spectra in each spaxel over the velocity range $-380$ to $0\kms$. The integrated line fluxes were then split about the velocity at which the two fitted [Fe II] 1.644 $\mu$m line components have the same flux density to form individual flux-ratio measurements for the high- and intermediate-velocity components. Spaxels were excluded where the relative uncertainty in the computed line ratio exceeded 20\% for the high-velocity component, and 50\% for the intermediate-velocity component. Therefore, line ratios for each component could only be obtained where the 1.533 $\mu$m [Fe II] emission line was detected with sufficient signal-to-noise ratio to satisfy this criterion. A flux-ratio map for the receding outflow will be presented in a future paper \citep{MCW13b}.

The electron number density of the approaching outflow high-velocity component is greatest $0\farcs 3$ to $0\farcs 5$ from the central star, with an average value of $\lesssim 4\times 10^4\textrm{ cm}^{-3}$. The electron density decreases to $\sim 10^{-4}\textrm{ cm}^{-3}$ within $0\farcs 8$ of the central star, and remains approximately constant to the edge of the observed field. There are no identifiable density enhancements at the positions of knots B and C. The electron number density of the intermediate-velocity component is more variable, between $\sim 10^{-3}\textrm{ cm}^{-3}$ to $\sim 10^{-4}\textrm{ cm}^{-3}$ within $0\farcs 9$ of the central star. Beyond that point, the signal-to-noise ratio of the [Fe II] 1.533 $\mu$m emission line is insufficient to form line ratios.

\begin{figure}
\centering
\includegraphics[width=\columnwidth]{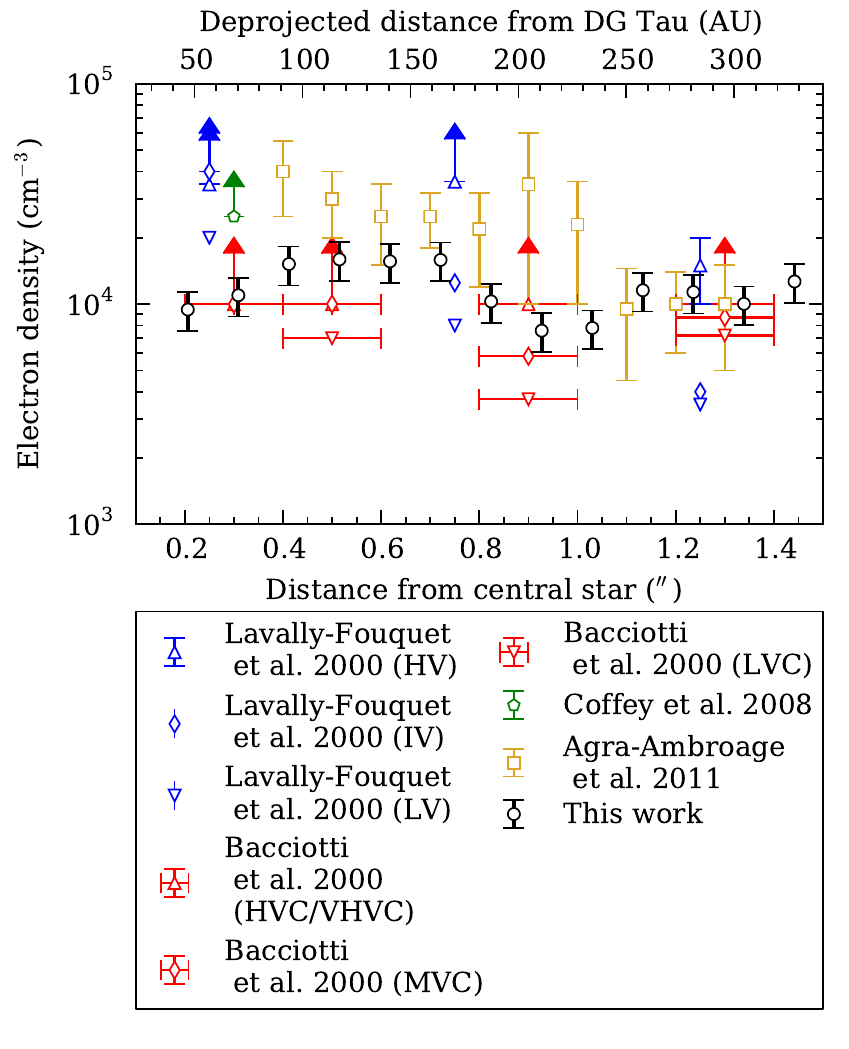}
\caption[Various electron density measurements of the DG Tau Jet]{Electron density measurements of the DG Tau approaching jet. Black circles show the electron density derived in this work for the DG Tau jet from the [Fe II] 1.533 $\mu$m/1.644 $\mu$m line ratio at each position along the outflow axis, averaged over all spaxels within $\pm 0\farcs 5$ of that axis in the perpendicular direction. All determinations of electron density from the literature are made using the optical line ratio technique developed by \citet{BE99}, except for those by \citet{A-Ae11}, which use the [Fe II] line ratio technique. Where provided, electron densities are quoted for high-velocity (HV), intermediate/medium-velocity (IV/MV) and intermediate-velocity (LV) components. Uncertainties are as quoted in the relevant reference, except for where they have been estimated from 2D maps of electron density \citep{Be00,CBP08}.}\label{fig:bluedenscompare}
\end{figure}

Our determination of the electron density of the approaching DG Tau jet (HVC) is compared with determinations from the literature in Fig.~\ref{fig:bluedenscompare}. 
We calculate an uncertainty-weighted average electron density at each position along the outflow axis from all spaxels within $\pm 0\farcs 5$ of the axis. Our results are in agreement with the previous determination of electron density from the [Fe II] line ratio by \citet{A-Ae11}, with slight discrepancies due to our different method of measuring the electron density of the jet component. 
Our results are also in reasonable agreement with electron density measurements of the outflow based on the [S II] 6716\AA /6731\AA\ line ratio using the BE99 technique \citep{BE99, LFCD00, Be00, CBP08}. 
Slight differences between electron densities derived from different spectral features are to be expected because the [S II] and [Fe II] lines arise in different regions of the cooling post-shock gas.

\section{Discussion}\label{sec:discuss}

In \S\ref{sec:obsdata} and \S\ref{sec:results}, we discussed our observations of the outflows from DG Tau at sub-arcsecond resolution, and with sufficient sensitivity to reveal their detailed structures. We have identified the approaching jet (high-velocity component), the blueshifted intermediate-velocity component, and the receding outflow. In this section, we discuss the origins and physical parameters for the blueshifted outflow components that can be inferred from these data. A detailed analysis of the nature of the receding outflow will be presented in a future paper \citep{MCW13b}.

\subsection{The Approaching Jet}\label{sec:D-bluejet}

We interpret the blueshifted [Fe II] 1.644 $\mu$m HVC emission to be from an approaching, high-velocity, well-collimated jet launched from the DG Tau star-disc system. We investigate the propagation of knots in the jet (\S\ref{sec:D-blueknots}), which leads us to identify a stationary recollimation shock in the jet channel (\S\ref{sec:D-recoll}). We use the properties of this shock to form estimates for the launch radii of the innermost streamlines of the jet (\S\ref{sec:D-launch}). We then proceed to calculate parameters of the jet downstream of this shock (\S\ref{sec:D-bluepower}), and investigate the cause of the changes in jet velocity along the outflow axis (\S\ref{sec:D-blueaccel}). Finally, we analyse our data for any indication of rotation in the DG Tau jet (\S\ref{sec:D-bluerot}).

\subsubsection{Knots}\label{sec:D-blueknots}

\begin{table*}
\caption{Knot positions in the approaching DG Tau jet, 2005--2009}\label{tab:knots050609}
\begin{tabular}{cccccccccc}
\hline
Knot & \multicolumn{1}{c}{} & \multicolumn{1}{c}{Positions} & \multicolumn{1}{c}{} & \multicolumn{2}{c}{Average} & \multicolumn{1}{c}{Centroid} & \multicolumn{2}{c}{Deprojected} & \multicolumn{1}{c}{Alternate} \\
     & \multicolumn{1}{c}{2005.87}   & \multicolumn{1}{c}{2006.98}    & \multicolumn{1}{c}{2009.88} & \multicolumn{2}{c}{proper motion\textsuperscript{a}} & \multicolumn{1}{c}{line velocity\textsuperscript{b}} & \multicolumn{2}{c}{velocties} & \multicolumn{1}{c}{designations} \\
     & & & & & & & \multicolumn{1}{c}{Proper} & \multicolumn{1}{c}{Radial} &  \\
     & \multicolumn{1}{c}{(\arcsec)}      & \multicolumn{1}{c}{(\arcsec)}       & \multicolumn{1}{c}{(\arcsec)} & \multicolumn{1}{c}{$(\arcsec\textrm{ yr}^{-1})$} & \multicolumn{1}{c}{$(\kms)$}  & \multicolumn{1}{c}{(2005.87,$\kms$)} & \multicolumn{2}{c}{($\kms$)} &  \\
\hline
A & 0.23$\pm$0.03 & 0.20$\pm$0.04 & 0.23$\pm$0.02 & 0   & 0 & --- & 0 & --- & --- \\
B & 0.40$\pm$0.03 & 0.60$\pm$0.02 & 1.07$\pm$0.02 & 0.17$\pm$0.01 & 113$\pm$7 & $\sim 180$ & $183\pm 11$ & $\sim 230$ & A5\textsuperscript{c} \\
C & 1.24$\pm$0.01 & ---           & ---           & --- & --- & $\sim 250$ & --- & $\sim 320$ & A3\textsuperscript{c} (?), A4\textsuperscript{c} (?) \\
\hline
\end{tabular}
\begin{flushleft}
Quoted uncertainties to the knot positions are the quadrature sum of the fitting errors to the star and knot positions. The fitting uncertainties for knot A are visual estimates.\\
Velocities are deprojected assuming an inclination of the jet axis to the line of sight of $38^\circ$ \citep{EM98}.\\
\textsuperscript{a}Fig.~\ref{fig:blueknotevolution}. \textsuperscript{b}Fig.~\ref{fig:bluevelfit}(b). \textsuperscript{c}\citet{A-Ae11}.
\end{flushleft}
\end{table*}

\begin{figure}
\centering
\includegraphics[width=\columnwidth]{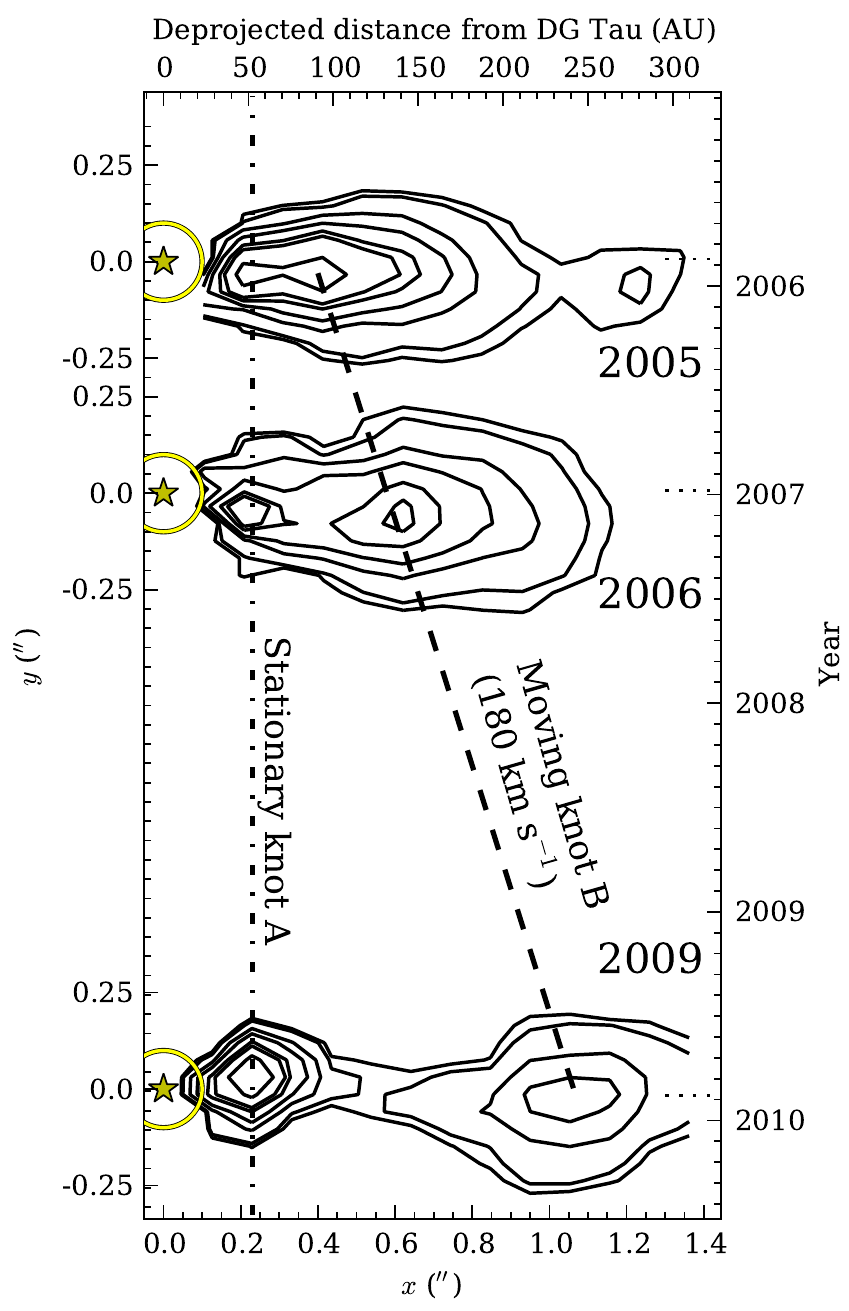}
\caption[Knots in the DG Tau Jet, 2005--2009]{Progression of knots in the approaching DG Tau outflow, 2005--2009. Shown is a contour plot of [Fe II] 1.644 $\mu$m line emission from the approaching DG Tau outflow at 2005.87, 2006.98 and 2009.88. Images are formed by integrating over the velocity range $-380\textrm{ to }0\kms$. Contours have levels of $[25,30,50,70,100,120,170]\times 10^{-15}\textrm{ erg cm}^{-2}\textrm{ s}^{-1}\textrm{ arcsec}^{-2}$. Short dotted lines represent the observation date of each epoch.}\label{fig:blueknotevolution}
\end{figure}

Three knots were observed in the DG Tau microjet (\S\ref{sec:knots}). Our unique multi-epoch data allow us to track the position of these knots over time, without the need to link disparate observations to form a knot evolution. The position of the knots as a function of time is given in Table \ref{tab:knots050609}, and shown in Fig.~\ref{fig:blueknotevolution}. The most remarkable finding is that knot A remains stationary over a period of $\sim 4$ years. We discuss the nature of this stationary feature in \S\ref{sec:D-recoll}.

\begin{figure}
\centering
\includegraphics[width=\columnwidth]{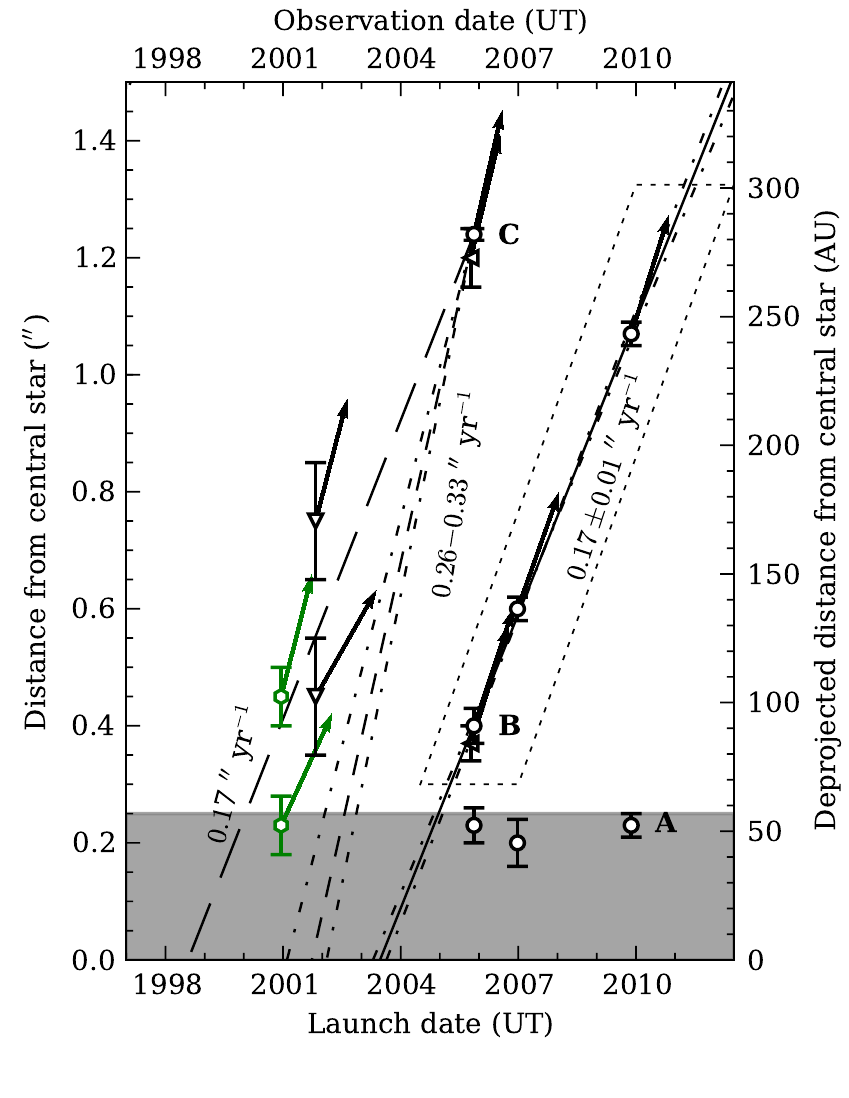}
\caption{Knot positions in the DG Tau approaching microjet at less than $1\farcs 5$ from the central star, plotted over the period 1997--2010, taken from multiple sources. The uncertainty in the knot positions have been visually estimated if uncertainties are not quoted in the relevant reference. Markers denote the source of the observations; colours denote the emission line(s) with which the observation was made. Where known, arrows denote the radial line velocity of the observed knot projected onto the plane of the sky. The solid line shows the linear fit made to the trajectory of knot B. The short-dashed line shows the trajectory of knot C assuming a proper motion of $0\farcs 30\textrm{ yr}^{-1}$ for that knot, corresponding to the radial velocity of that knot projected onto the sky using a jet inclination of 38$^\circ$. Dot-dashed lines show the uncertainties in these trajectories: for knot B, this is the fitting uncertainty; for knot C, this results from a $\pm 3.5^\circ$ variation in the jet inclination (\S\ref{sec:jettraj}). The long-dashed line shows the trajectory of knot C assuming a proper motion of $0\farcs 17$ for that knot, as for knot B. The grey area denotes the region $\leq 0\farcs 25$ from the central star, where knot observations are excluded from fitting. Knot observations used for fitting the trajectory of knot B are grouped by the dotted parallelogram.
\newline References. \emph{Hexagons: } \citet{Te02}. \emph{Down-pointing triangles:} \citet{Pe03b}. \emph{Left-pointing triangles:} \citet{A-Ae11}. \emph{Circles:} This work.
\newline \emph{Green:} He I 1.0830 $\mu$m. \emph{Black:} [Fe II] 1.644 $\mu$m.}\label{fig:blueknots05}
\end{figure}

We were able to track the progression of knot B over this interval. The knot moves at a constant speed of $0\farcs 17\pm 0\farcs 01\textrm{ yr}^{-1}$ along the jet channel, which implies a knot launch date of $2003.5\pm 0.2$ by linear extrapolation\footnote{This extrapolation includes the position of knot B/A5 quoted by \citet{A-Ae11}.} (Fig.~\ref{fig:blueknots05}). This speed is slower than that of knots previously observed in the DG Tau jet \citep[e.g., $0\farcs 29\textrm{ yr}^{-1}$,][]{De00}, and slower than the knot proper motions of $0\farcs 27\textrm{--}0 \farcs 34\textrm{ yr}^{-1}$ suggested by \citet{A-Ae11}. 

The presently-favoured model for the formation of moving jet knots is intrinsic variability in the jet velocity \citep[e.g.,][]{Rae90}. In this scenario, as the jet velocity oscillates, faster regions of the jet catch up to slower-moving regions, forming shocked internal working surfaces which appear as jet knots. Our data show evidence of such velocity variations in the jet (\S\ref{sec:D-blueaccel}), in agreement with previous studies \citep[e.g.,][]{Pe03b}. A basic prediction of this theory is that the proper motion and radial velocity of the shocked material in the knots should be two projections of the same knot velocity \citep{Rae90}. This does not appear to be the case for knot B (Table \ref{tab:knots050609}) if we assume a constant jet inclination and therefore adopt a jet inclination of $38^\circ$ as determined on scales of $\sim 10\arcsec$ \citep{EM98}. However, the jet ridgeline is not a straight line, and the jet inclination may therefore vary locally by $\sim 3\textrm{--}4^\circ$ (\S\ref{sec:jettraj}). Taking a local jet inclination of $34.5^\circ$ at the location of knot B reconciles the proper motion and radial velocity when deprojected. Therefore, we conclude that knot B could have been formed by intrinsic velocity variations in the jet.

The knot periodicity of DG Tau has been studied by previous authors, but has remained unclear. \citet{Pe03b} determined a knot ejection period of $\sim 5\textrm{ yr}$, which was revised downwards to $2.5\textrm{ yr}$ by \citet{A-Ae11}. However, \citet{Re12} used a different interpretation of knot motions to claim a $5\textrm{ yr}$ ejection period. Our observations support the notion that the knot ejection interval in DG Tau varies. We offer the following arguments in favour of this interpretation. First, we note the absence of a new moving knot in our 2009 data (Figs.~\ref{fig:blueknotevolution}, \ref{fig:blueknots05}). We would expect to observe a new knot somewhere between knots A and B at this epoch if there were a knot ejection every $2.5\textrm{ yr}$, hence we exclude the proposition that a new knot was launched $2.5\textrm{ yr}$ after knot B.

Second, we consider the other moving knot in our data, knot C. This knot only appears in our 2005 data, having moved out of the NIFS field by 2006.98 (Fig.~\ref{fig:blueknotevolution}). We do not attempt to link this knot directly to previous observations, due to the inherent uncertainty in doing so (see below). We instead assign knot C two possible proper motions, and examine the implications of each scenario. First, we presume that the radial velocity of knot C ($\sim 250\kms$, Table \ref{tab:knots050609}) represents a projection of the true knot velocity. Allowing for a variation of $\pm 3.5^\circ$ in the canonical jet inclination of $38^\circ$ \citep[\S\ref{sec:jettraj};][]{EM98}, this gives a proper motion for knot C of $0\farcs 26\textrm{--}0 \farcs 33\textrm{ yr}^{-1}$. This is consistent with the interpretation of \citet{A-Ae11}, and yields a knot launch date of $2001.7^{+0.2}_{-0.6}$, consistent with a $2.5\textrm{ yr}$ ejection period. Second, we assign knot C a proper motion of $0\farcs 17\textrm{ yr}^{-1}$, which is consistent with the proper motion of knot B. This is significantly slower than the knot velocity implied by the radial velocity of knot C; however, we note that the knots in Herbig-Haro objects often show discrepancies between their proper motion and radial velocity \citep{EM92,EM94}. This knot trajectory passes through the cluster of knot observations reported by \citet{Te02} and \citet{Pe03b} (Fig.~\ref{fig:blueknots05}). This proper motion gives a launch date of 1998.6 for knot C, which would imply a knot launch period of $\sim 5\textrm{ yr}$.

In light of the above complications, we leave the knot ejection interval in DG Tau, and the true knot velocity of knot C, as open questions. We have not attempted to directly link our knot observations with those from the literature. This is because, with the exception of the fast knot detected by \citet{De00} mentioned above, most DG Tau jet knots reported in the literature are single observations made in different emission lines, and using different instruments. This means that disparate observations need to be linked to form an interpretation of the knot ejection history. We prefer to await further, consistent multi-epoch data of the DG Tau jet in order to attempt to draw a final conclusion on the knot ejection interval of this object. Indeed, in light of the large difference between the knot proper motions observed by us and \citet{De00}, we suggest there is significant variability in the ejection interval, and in the knot ejection velocity. We are intrigued to see if there will be a repeat of the fast knot reported by \citet{De00} at a later date. However, if the knot ejection interval and velocity are reasonably constant with a a $5\textrm{ yr}$ ejection period, we predict that a new jet knot should have been launched from the position of the central star in mid-2008, and would have become visible beyond the stationary recollimation shock in approximately mid-2010. There is currently no available data with which to test this hypothesis.

\subsubsection{The Recollimation Shock}\label{sec:D-recoll}

We interpret knot A in the approaching jet as a stationary recollimation shock. Stationary [O I] 6300 \AA\ emission in the region of this feature has been detected previously by \citet{Le97}, $\sim 0\farcs 15\approx 24\textrm{ AU}$ from the central star. Stationary soft X-ray emission has been observed in the DG Tau jet, centred $\sim 0\farcs 14\textrm{--}0\farcs 21\approx 32\textrm{--}48\textrm{ AU}$ from the central star \citep{Ge05,Ge08,Ge11,SS08,GML09}. Stationary far-ultraviolet C IV emission is observed slightly further along the jet, centred $0\farcs 2\approx 46\textrm{ AU}$ from the central star \citep{Sce13}. The temperature of the X-ray emitting material is estimated to be $\gtrsim 3\times 10^{6}\textrm{ K}$ \citep{Ge08,GML09}, whilst the emissivity of C IV strongly peaks at temperatures of $10^5\textrm{ K}$ \citep{Sce13}. We interpret this as being indicative of an extended post-shock cooling region, where a recollimation shock occurs $\sim 25\textrm{ AU}$ from the central star, and material then cools as it progress downstream on the scale of a cooling length \citep[e.g.,][]{Fe14}.

In classical hydrodynamic jet theory, recollimation shocks appear when a jet emerging from a nozzle is under- or over-expanded, and undergoes lateral expansion and/or contraction to attain pressure equilibrium with the ambient medium. In the context of magnetocentrifugally driven jets and winds, recollimation of outflows into stationary shocks above the disc is due to the magnetic field acting like a nozzle. Recollimation shocks occur naturally in magnetocentrifugal outflows with terminal poloidal velocities $\gtrsim 2$ times the fast magnetosonic speed in the outflowing material \citep{GdCP93}. In this scenario, once the flow expands to a critical radius, the magnetic tension acting inwards towards the outflow axis exceeds the centrifugal force acting outwards, and the jet begins to recollimate into a stationary shock \citep{BP82,CL94}. Such recollimation shocks have been explored in the theoretical literature, and are predicted to occur tens of AU above the circumstellar disc for reasonable YSO accretion rates and disc parameters \citep{PP92,GdCP93,OP93,GdCV01,FC04}, in agreement with observations of the stationary feature in the approaching DG Tau jet. Finally, \citet{Aie13} utilised \emph{e-MERLIN} data to measure the opening angle of the DG Tau jet to be $86^\circ$, implying that collimation must occur somewhere $\backsimeq 50\textrm{ AU}$ along the jet channel. This is in excellent agreement with the observed position of the stationary feature in the DG Tau jet.

Previous analyses of the stationary soft X-ray emission in the DG Tau jet have concluded that the mass flux through the X-ray emitting region is $\sim\textrm{ a few}\times 10^{-11}\Msol$ \citep{SS08,GML09}, which is two orders of magnitude less than the mass flux seen in the NIR/optical (\S\ref{sec:D-blueaccel}). However, the geometry used by \citet{SS08} and \citet{GML09} to compute the X-ray mass flux may result in an underestimation. Those authors used a cylindrical geometry of height $d_\textrm{cool}$ (i.e.,~the adiabatic cooling length) and radius $R$ to describe the X-ray emitting region. However, \citet{Be11} generated a numerical simulation of a YSO jet recollimation shock\footnote{This simulation involves the launching of a jet with a uniform cross-jet velocity profile, which is forced to recollimate after passing through a nozzle.} to investigate the stationary X-ray emission in the outflow from L1551 IRS5\footnote{The large-scale outflow this object drives is HH 154. The soft X-ray knot in the outflow is located $0\farcs 5\textrm{--}1\farcs 0$ from the outflow source \citep{Be11}.}, which shows that such shocks take on an inverted-cone structure. Repeating the calculation of \citet{SS08} using an inverted cone of height $d_\textrm{cool}$ and radius $R$ has two effects on the result. Firstly, the volume of the emitting region is decreased by a factor of 3. Second, the area through which mass enters the X-ray emitting region is increased by a factor of $\sqrt{1+(d_\textrm{cool}/R)^2}$, where $d_\textrm{cool}/R\sim 4$ from the simulation of \citeauthor{Be11} This results in a mass flux $\sim 12$ times higher than that reported by \citet{SS08} and \citet{GML09}, which for DG Tau increases the X-ray mass flux to $\sim\textrm{ few}\times 10^{-10}\Msol$.

There remains a discrepancy of at least an order of magnitude between the X-ray and optical/NIR-derived mass fluxes. This has led several authors to suggest that there must be an inner, very fast component of the DG Tau approaching outflow, not visible at other wavelengths and perhaps of stellar or magnetospheric origin, nested within the optical/NIR high-velocity outflow component \citep[e.g.,][]{GML09,Fe14}. The mass-flux discrepancy may be explained by considering the geometry and emission characteristics of a diamond recollimation shock. The results of the simulation of \citet{Be11} show that, whilst the \emph{entire} jet is shocked to a temperature of $\gtrsim 10^{6}\textrm{ K}$ around the recollimation shock, only the small central core of the diamond structure significantly emits in X-rays \citep[][fig.~4 therein, right-hand panels]{Be11}. The balance of the jet material is focused around the central emission peak by the diamond shock structure, and does not achieve the pressure necessary to strongly emit in X-rays. This neatly explains the discrepancy between the mass flow rates of the soft X-ray source and optical/NIR flow in DG Tau. Hydrodynamic simulation of the recollimation shock in DG Tau is required in order to quantify the expected mass flux through the X-ray emitting region.

There are alternate explanations for the presence of this stationary feature. It has been suggested that stationary knots in the outflows from massive protostars may be the result of the stellar wind bouncing off the walls of a cleared jet channel and recollimating above the stellar surface \citep{Pae09}. As mentioned above, there may also be another, unresolved central outflow component, possibly launched from the magnetosphere of the star, that may be the cause of this hot X-ray emission \citep[e.g.][]{FDC06}. Our interpretation has the advantage that it does not require invoking an as-yet undetected outflow component. Regardless of its origin, the presence of such a bright, strong, hot shock directly in the jet channel must affect the jet material that passes through/by it.

The implications of the presence of stationary recollimation shocks on the study of YSO jets are profound. The shock will modify the flow parameters downstream of its position. Therefore, extreme care and caution is required when attempting to link parameters in the outflow beyond the stationary shock, such as terminal velocities, to a specific launch radius (\S\ref{sec:D-launch}). Passage through such a strong shock will create turbulence in the jet, and may remove any jet rotational signature (\S\ref{sec:D-bluerot}). 
We now proceed to investigate each of these in detail.

\subsubsection{Innermost Jet Streamlines: Terminal Velocity and Launch Radius}\label{sec:D-launch}

Determining the radii at which protostellar outflows are launched is one of the major goals of studies such as ours. Determination of launch radius is crucial information for determining the outflow launch mechanism. A constraint on the launch radius of protostellar outflows can be arrived at from measurements of the poloidal and toroidal jet velocities at some distance from the central star under a steady, magnetocentrifugal acceleration model \citep{Ae03}. \citet{FDC06} provide a diagnostic diagram to this end, for various forms of MHD wind acceleration.  
However, this method must be applied with caution to DG Tau. We must account for the presence of the strong recollimation shock in the outflow channel (\S\ref{sec:D-recoll}). Furthermore, we find no evidence for rotation in the DG Tau jet (\S\ref{sec:D-bluerot}).
Therefore, we proceed to make an estimate of the launch radius of the innermost streamlines of the DG Tau jet including the observed properties of the recollimation shock, assuming that these streamlines are launched by an MHD disc wind. We consider pressure-driven stellar winds at the end of this section.

We estimate the launch radius of the innermost radii of the DG Tau jet as follows. For magnetocentrifugal, axisymmetric winds, the specific energy of the flow, which is constant along field lines, can be expressed as
\begin{equation}
E=\frac{1}{2}\left(v_p^2+v_\phi^2\right)+\phi+h+\Omega_0\left(\Omega_0 r_\textrm{A}^2-\Omega r^2\right)
\end{equation}
\citep[e.g.,][]{KP00,KS11}, where $\phi$ is the gravitational potential, $h$ is the specific enthalpy, $r_\textrm{A}$ is the Alfv\'{e}n radius, i.e.~the radius at which the outflow velocity equals the Alfv\'{e}n speed, $r$ is the radial distance from the central star, $v_\textrm{p}$ and $v_\phi$ are the flow poloidal and azimuthal velocity components, respectively, and $\Omega$ is the angular velocity; subscript zero denotes values at the flow footpoint. For dynamically cold flows of gas, the enthalpy term can be neglected, and the gravitational potential is usually considered unimportant far from the disc. Further assuming that $E\approx v_{p,\infty}^2/2$ as $r\rightarrow\infty$, where $v_{p,\infty}$ is the flow poloidal velocity at large distances, and that $(r_\textrm{A}/r_0)^2\gg 1$, the terminal flow poloidal velocity may be written as
\begin{equation}\label{eq:velinf-init}
v_{\textrm{p},\infty} \simeq \sqrt{2} \Omega_\textrm{K} r_\textrm{A}\textrm{.}
\end{equation}
This equation can be obtained from equation (8) of \citet{FDC06} by neglecting their $\beta$ term, which encompasses all pressure effects, and assuming that their parameter $\lambda_\phi=rv_\phi / \Omega_0r_0^2\gg 3/2$. The Keplerian angular velocity, $\Omega_\textrm{K}$, at the disc launch radius of the wind, $r_0$, is given by
\begin{equation}
\Omega_\textrm{K}=\frac{v_\textrm{K}}{r_0}=\frac{1}{r_0}\left(\frac{GM_\star}{r_0}\right)^{1/2}\textrm{,}
\end{equation}
so equation (\ref{eq:velinf-init}) becomes
\begin{equation}\label{eq:velinf-final}
v_{\textrm{p},\infty}\simeq \sqrt{2}v_\textrm{K}\frac{r_\textrm{A}}{r_0} = \sqrt{2}\left(\frac{GM_\star}{r_0}\right)^{1/2}\left(\frac{r_\textrm{A}}{r_0}\right)\textrm{.}
\end{equation}
A stellar mass for DG Tau of $M_\star = 0.67 \textrm{ }M_\odot$ is adopted \citep{HEG95}. Then, for convenience, equation (\ref{eq:velinf-final}) can be expressed as
\begin{equation}\label{eq:velinf-units}
v_{\textrm{p},\infty}\simeq 109\kms \left(\frac{r_0}{0.1\textrm{ AU}}\right)^{-1/2}\left(\frac{r_\textrm{A}}{r_0}\right)\textrm{.}
\end{equation}

A wide range of values are both observationally justified and theoretically possible for the magnetic lever arm parameter, $\lambda=(r_\textrm{A}/r_0)^2$. \citet{CF00} calculated steady MHD wind solutions for $\lambda$ exceeding $\sim 2 \Rightarrow (r_\textrm{A}/r_0)\gtrsim 1.4$. In the analysis of the launch radii of various protostellar outflows by \citet{FDC06}, the observationally-inferred magnetic lever arm $\lambda_\phi$ for high-velocity outflows is in the range $4\textrm{--}16$. Given that the observational estimate $\lambda_\phi$ may underestimate the true $\lambda$ due to the sampling of multiple magnetic surfaces in the jet \citep{FDC06}, we adopt a range of $4 \leq \lambda \leq 20$, which leads to $2 \leq r_\textrm{A}/r_0 \lesssim 4.5$, as an illustrative parameter range for YSO jets. We also note the typical observation that the ratio of mass outflow rate to mass accretion rate, $\dot{M}_\textrm{out}/\dot{M}_\textrm{acc}\sim 0.1$, implies $\lambda\sim\textrm{ a few to }10$ assuming that the rate at which angular momentum is lost by the accreting matter ($\dot{M}_\textrm{acc}r_0^2/\Omega_0$) equals the rate of angular momentum transport by the wind \citep[$\dot{M}_\textrm{out}r_A^2/\Omega_0$; see][]{C07b}.

\begin{figure}
\centering
\includegraphics[width=\columnwidth]{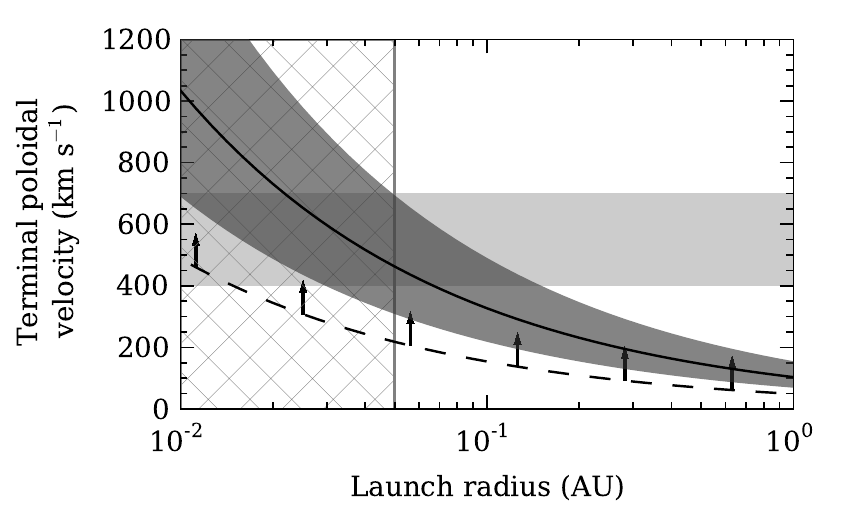} 
\caption{Predicted DG Tau asymptotic poloidal jet velocities as a function of launch radius, assuming an MHD disc wind. The solid line denotes the solution for a stellar mass of $M_\star=0.67\textrm{ }M_\odot$ \citep{HEG95} and $r_\textrm{A}/r_0 =3$ \citep[e.g.,][]{KS11}. The dark grey region shows the range of solutions for $2 \leq r_\textrm{A}/r_0 \leq 4.5$ \citep{FDC06}; the dashed line shows the limit of solutions for $1.4 \leq r_\textrm{A}/r_0$ \citep{CF00}. The light grey horizontal bar represent the range of possible terminal velocities for the DG Tau jet, based on analysis of the stationary recollimation shock (\S\ref{sec:D-recoll}). The grey hatching denotes a launch radius of less than $0.05\textrm{ AU}$, where the jet could be launched via the X-wind mechanism.\label{fig:bluejetlaunch}}
\end{figure}

The presence of the stationary recollimation shock (\S\ref{sec:D-recoll}) must be taken into account when determining the terminal poloidal velocity, $v_{\textrm{p},\infty}$, of the jet. Under standard theories of magnetocentrifugal acceleration, terminal velocity is reached beyond the fast magnetosonic point in the outflow, which is predicted to be a few tens of AU above the circumstellar disc surface at most \citep[e.g.,][]{GdCP93,C07}. Most authors assume acceleration largely ceases beyond this point, and the jet then flows ballistically. However, a stationary recollimation shock will slow the jet material, so that the jet velocity observed immediately beyond knot A will not be indicative of the magnetocentrifugal terminal velocity. X-ray observations of the stationary knot suggest a shock velocity of $400-600\kms$, based on an inferred shock temperature of \mbox{3--4} MK \citep{Ge08,SS08,GML09}. Further observations have indicated that this shock velocity may be as high as $700\kms$ (M.~G\"{u}del, private communication). For a stationary shock, the shock velocity is equal to the pre-shock gas velocity. Therefore, the innermost streamlines of the DG Tau jet must be accelerated to a terminal poloidal velocity of $v_{\textrm{p},\infty}\sim 400\textrm{--}700\kms$ in order to form the observed shock. This is a significantly higher poloidal velocity than used by previous authors to determine the launch radius of the DG Tau jet \citep{Ce07}. Such pre-shock velocities for the jet core were proposed for DG Tau by \citet{GML09}. However, the implied launch radius of such a jet was not considered therein.

The terminal poloidal velocity of the innermost streamlines of DG Tau jet, equation (\ref{eq:velinf-units}), is plotted as a function of launch radius for DG Tau in Fig.~\ref{fig:bluejetlaunch}, for a range of magnetic level arm values. For a terminal jet velocity of $v_{\textrm{p},\infty}\approx 400\textrm{--}700\kms$, we determine a jet launch radius of $0.01\textrm{--}0.15\textrm{ AU}$ for the innermost jet streamlines, using $2 \leq r_\textrm{A}/r_0 \lesssim 4.5$ \citep[e.g.,][table 1 therein]{FDC06}. Using the canonical value $r_\textrm{A}/r_0 =3$ gives a launch radius range of $0.02\textrm{--}0.07\textrm{ AU}$. The constraint $r_\textrm{A}/r_0 \gtrsim 1.4$ yields a minimum launch radius of $0.005\textrm{ AU}$. We note that outer jet streamlines that do not radiate in X-rays may be launched from larger radii. Previous estimates of the launch radius of the jet have been in the range $\lesssim 0.1\textrm{ AU}$ \citep{Ae03} to $0.3\textrm{--}0.5\textrm{ AU}$ \citep{Ce07}. The smaller launch radius calculated here is a direct result of using a significantly higher jet terminal poloidal velocity. For comparison, had we inferred a terminal poloidal velocity of $\sim 215\kms$ from the approaching jet velocity after the stationary shock, we would have calculated a launch radius of 0.23 AU for $r_\textrm{A}/r_0 =3$. A jet launched from such a radius would unequivocally be interpreted as originating from a disc wind.

We are unable to exclude the possibility that the innermost streamlines in the DG Tau jet originate from a magnetospheric wind such as the X-wind \citep{Se00}. The stellar radius of DG Tau has been determined previously to be $2.5R_\odot\approx 0.01\textrm{ AU}$ \citep{Ge07}, so that launch points within a few stellar radii of the central star, characteristic of an X-wind, are possible. However, the circumstellar disc may also approach this close to the central star \citep[e.g.,][]{PP92,GdCP93}, so that disc wind contribution to this fast outflow is feasible. Indeed, an MHD disc wind launched from a radius of five stellar radii, approximately 0.05 AU for DG Tau, would most readily explain the high ejection-accretion efficiencies generally observed in YSOs \citep{C07b}. Finally, we note that a pressure-driven stellar wind with a ratio of thermal to magnetic pressure, $\beta$, between 5.2 and 11.8 could also produce a $400\textrm{--}700\kms$ wind in DG Tau \citep[assuming a magnetic level arm parameter $\lambda<200$;][]{FDC06}.

\subsubsection{Jet Parameters}\label{sec:D-bluepower}

The parameters of the approaching jet have been computed based on the HVC line fits (Fig.~\ref{fig:bluevelfit}(a,b)) and density estimates (Fig.~\ref{fig:densrat}). Jet parameters are essential in order to compare these observational results with numerical simulations of the DG Tau outflows. Determining the jet mass flux is also useful as an input for modelling the receding outflow \citep{MCW13b}.

\begin{figure}
\centering
\includegraphics[width=\columnwidth]{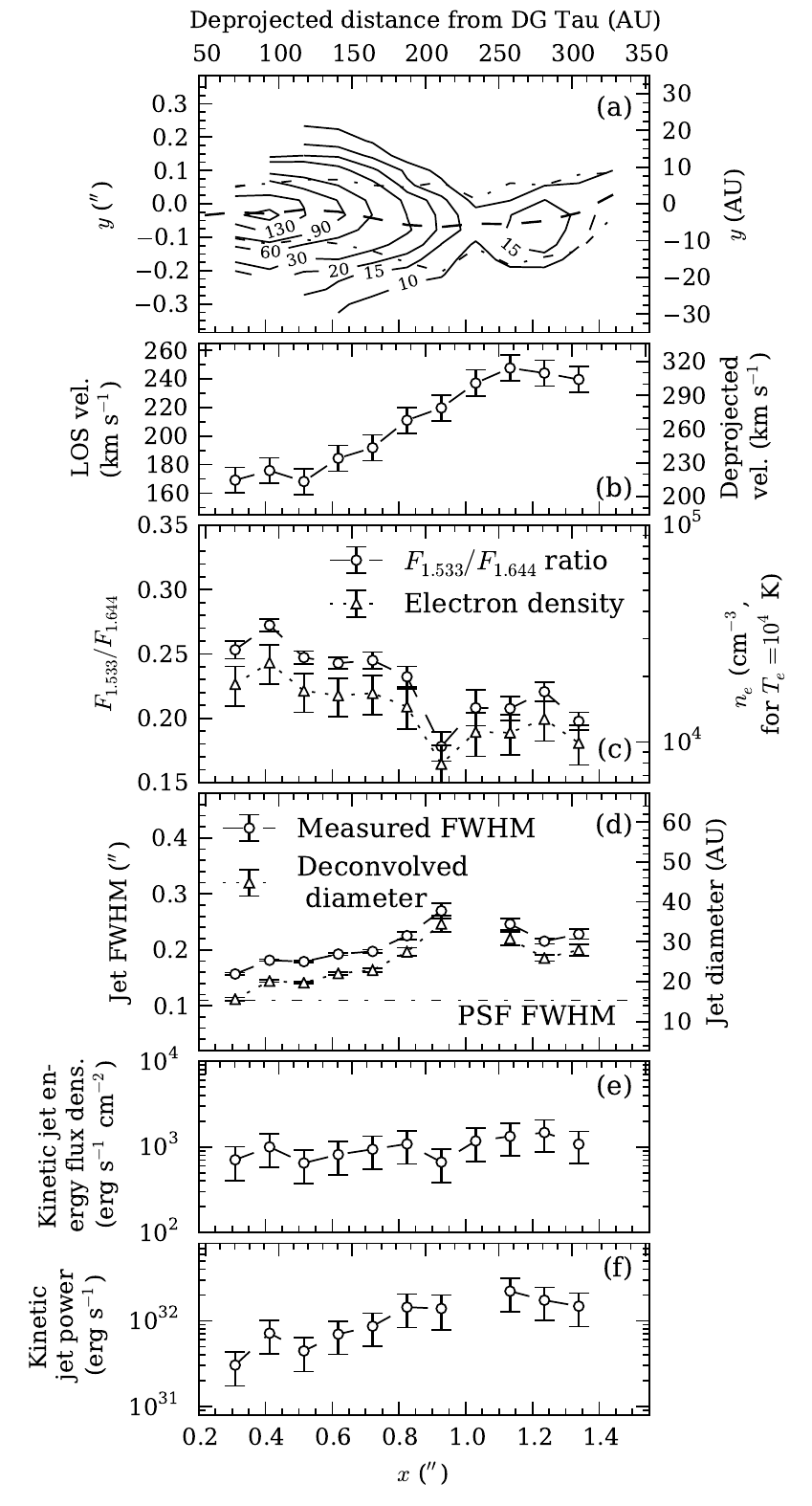}
\caption[Blueshifted HVC jet parameters]{Derived parameters for the approaching DG Tau jet. (a) Contours of [Fe II] 1.644 $\mu$m blueshifted HVC emission, in units of $10^{-15}\textrm{ erg cm}^{-2}\textrm{ s}^{-1}\textrm{ arcsec}^{-2}$. The unlabelled contour is at $160\times 10^{-15}\textrm{ erg cm}^{-2}\textrm{ s}^{-1}\textrm{ arcsec}^{-2}$. (b) Line-of-sight and deprojected [Fe II] $1.644\textrm{ }\mu\textrm{m}$ HVC line velocity along the jet ridgeline. The errorbars are the quadrature sum of the fitting and stellar velocity uncertainties. (c) [Fe II] $1.533\textrm{ }\mu\textrm{m}/1.644\textrm{ }\mu\textrm{m}$ ratio and electron density along the jet ridgeline. The electron density is calculated for an electron temperature of $10^4\textrm{ K}$. (d) Observed and deconvolved jet FWHM of the blueshifted HVC. The PSF FWHM is shown by the dot-dashed line. (e) Computed jet kinetic energy flux density along the jet ridgeline. (f) Computed jet kinetic power, $L_\textrm{jet}$, along the jet ridgeline.}\label{fig:bluepower}
\end{figure}

The derived parameters of the approaching DG Tau jet are shown in Fig.~\ref{fig:bluepower}. From top to bottom, the panels correspond to integrated [Fe II] 1.644 $\mu$m emission-line intensity, velocity, density, diameter, kinetic energy flux density, and kinetic power. Each measurement is derived from the spaxel that is closest to the jet ridgeline at each position along the outflow axis. Axial distances and velocities are deprojected using a jet inclination to the line of sight of 38$^\circ$. This inclination was determined by comparing the radial and proper motions of the bow shock at the head of the HH 158 outflow \citep{EM98}. The 1.533 $\mu$m/1.644 $\mu$m line ratio is converted to electron density using the formula presented by \citeauthor{A-Ae11} \citep[2011; based on][]{Pes03}, which they claim has an intrinsic accuracy of 20\%. This calculation is performed for an electron temperature of $10^4\textrm{ K}$, as estimated for the DG Tau jet by \citet{B02} through ratios of optical lines \citep{BE99}. The jet diameter $D_\textrm{jet}$ is estimated by forming Gaussian fits to the HVC integrated intensity image (Fig.~\ref{fig:bluepower}(a)) transverse to the jet direction, and then approximately deconvolving this width from the PSF via the formula $D_\textrm{jet}^2=
\textrm{FWHM}_\textrm{obs}^2-\textrm{FWHM}_\textrm{PSF}^2$.

In order to calculate the kinetic jet energy flux, it is necessary to determine the jet density, $\rho_\textrm{jet}$. To accomplish this, the electron density, $n_e$, determined from the [Fe II] line ratio was converted to jet density using the formula $\rho_\textrm{jet}=n_\textrm{H}m_\textrm{H}\mu$, where $\mu=1.4$ for a typical gas composition of 90\% hydrogen and 10\% helium. The hydrogen number density, $n_\textrm{H}$, is given by the ratio of the electron density and the ionisation fraction, $\chi_e$, which is taken to be $\chi_e=0.3\pm 0.1$, as determined for the high-velocity components of the DG Tau jet by \citet{B02}, and later refined by \citet{Me14}, from ratios of optical lines \citep{BE99}. The jet kinetic energy flux density is then calculated via the formula $F_\textrm{E}=(1/2)\rho_\textrm{jet}v_\textrm{jet}^3$. Finally, multiplying the jet kinetic energy flux density by the jet cross-sectional area, estimated as $A_\textrm{jet}=\pi (D_\textrm{jet}/2)^2$, gives the jet kinetic power. 

The approaching jet from DG Tau is observed to accelerate from a deprojected velocity $\sim 215\kms$ to $\sim 315\kms$ over the region $0\farcs 5\textrm{--}1\farcs 15\approx 115\textrm{--}260\textrm{ AU}$ from the central star (Fig.~\ref{fig:bluepower}(b)).  This corresponds to a region of steadily increasing jet diameter, from $\sim 19\textrm{ AU}$ to $\sim 28\textrm{ AU}$ (Fig.~\ref{fig:bluepower}(b)). The jet kinetic power increases over this region, from $(4.4\pm 1.9)\times 10^{31}\textrm{ erg s}^{-1}$ to $(2.2\pm 0.9)\times 10^{32}\textrm{ erg s}^{-1}$. The jet acceleration and related increase in jet kinetic power are discussed further is \S\ref{sec:D-blueaccel}.

\begin{figure}
\centering
\includegraphics[width=\columnwidth]{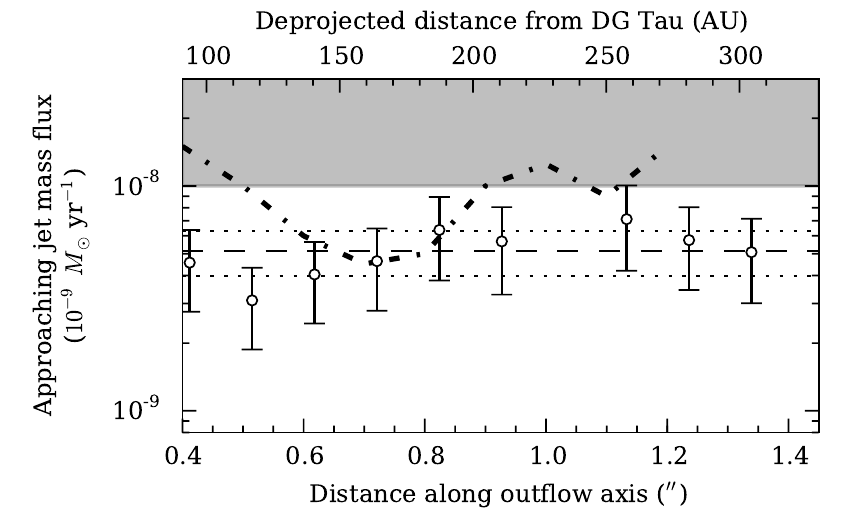}
\caption{Jet mass flux in the approaching jet of DG Tau. The circles and error bars show the mass flux computed from the measured physical parameters of the jet, and the associated uncertainties. The dashed line represents the average value of all data points, and the dotted lines show the standard deviation of the measurements. The jet mass flux determined by \citet{A-Ae11} via a similar method is shown as a thick dot-dashed line. The jet mass flux determined by \citet{Le13} from VLA data is represented by the grey shaded region.}\label{fig:bluemassflux}
\end{figure}

The approaching jet mass flux, determined by the formula $\dot{M}=\rho_\textrm{jet}v_\textrm{jet}A_\textrm{jet}$, is shown in Fig.~\ref{fig:bluemassflux}. The jet mass flux is constant within measurement errors, with an average value of $(5.1\pm 1.2)\times 10^{-9}\textrm{ }M_\odot\textrm{ yr}^{-1}$. Our measurements typically agree to within 1$\sigma$ with the measurements made by \citet{A-Ae11} using a similar technique, with discrepancies most likely due to our differing methods of determining the electron density of the jet. Our jet mass flux is also consistent with that determined by \citet{Me14}, $(8\pm 4)\times{10}^{-9}\textrm{ }M_\odot\textrm{ yr}^{-1}$. Both our mass flux determination and that of \citet{A-Ae11} are lower than previous estimates from VLA data of $\dot{M}\sim 1\textrm{--}5\times 10^8\textrm{ }M_\odot\textrm{ yr}^{-1}$ \citep{Le13}. However, as noted by those authors, the uncertainties in estimating this quantity makes detailed comparison difficult. We conclude that the DG Tau jet has a constant mass flux within $\sim 350\textrm{ AU}$ of the central star within our measurement uncertainties.

\subsubsection{Jet Velocity Variability}\label{sec:D-blueaccel}

\begin{figure}
\centering
\includegraphics[width=\columnwidth]{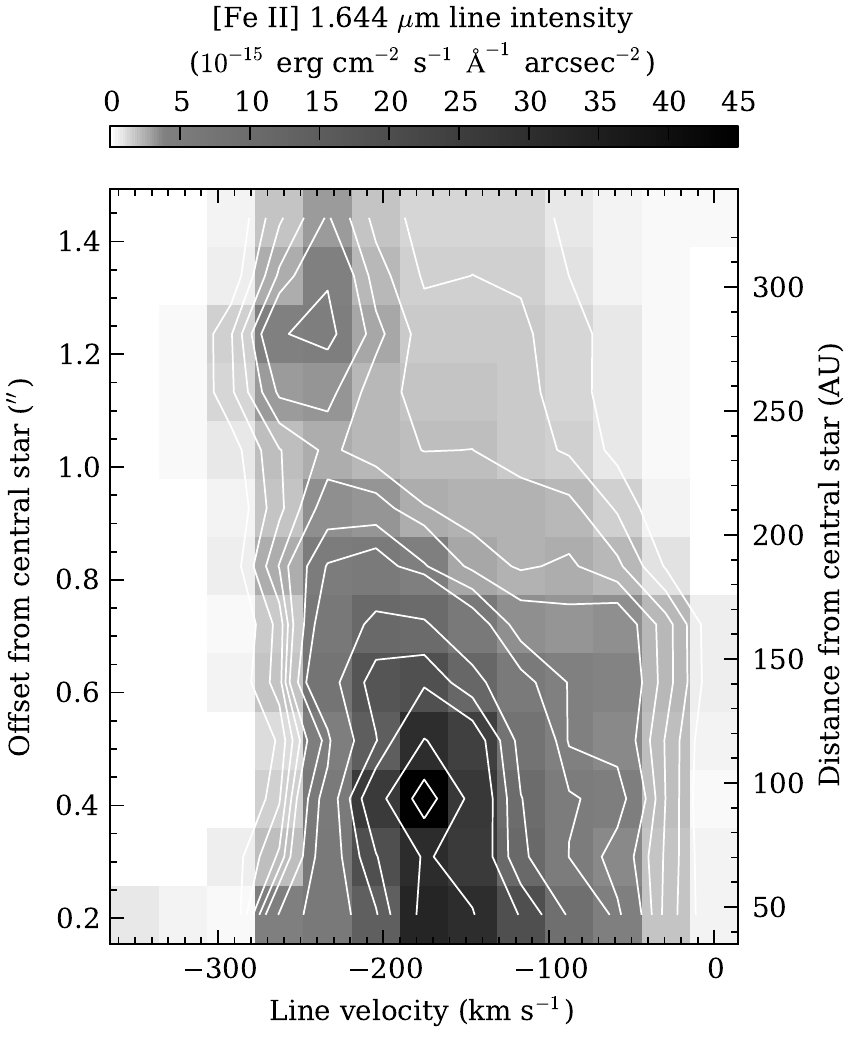}
\caption{Position-velocity diagram of the blueshifted outflow from DG Tau. At each downstream position, the spaxel containing the jet ridgeline is dispersed. Contours are plotted at levels of $[1$, $1.5$, $\ldots$, $3$, $4$, $5$, $10$, $15$, $20$, $30$, $40]\times 10^{-15}\textrm{ erg cm}^{-2}\textrm{ s}^{-1}\textrm{ \AA}^{-1}\textrm{ arcsec}^{-2}$.}\label{fig:bluePV}
\end{figure}

Hydromagnetic winds are initially accelerated via magnetocentrifugal processes, which are efficient up to approximately the Alfv\'{e}n critical surface \citep[e.g.,][]{BP82}. This surface is expected to be located within at most a few tens of AU of the central star \citep[e.g.,][]{GdCP93,C07b}. However, our data show a clear increase in velocity in the approaching jet over the region $\sim 115\textrm{--}260\textrm{ AU}$ from the central star, well beyond the predicted location of the Alfv\'{e}n surface. This trend can also be seen in a position-velocity diagram of the approaching outflow, formed along the jet axis (Fig.~\ref{fig:bluePV}). The increase in velocity is smooth, with no sudden velocity changes. This acceleration has been observed in previous studies of the approaching jet from DG Tau \citep{Be00,Te04,A-Ae11}, but has not been definitively explained. It is possible that this ``acceleration'' is a stroboscopic effect when observing a jet with intrinsic velocity variations, as suggested by the observations of \citet{Pe03b}. We discuss other possible causes of this apparent acceleration below.

Purely hydrodynamic pressure-driven acceleration is not possible in the DG Tau jet for the following reasons. Acceleration in a jet may be driven by thermal pressure coupled with expansion. In the adiabatic case, this process is governed by the Bernoulli equation,
\begin{equation}\label{eq:Bern}
\frac{1}{2}v^2 + h = \textrm{const.,}
\end{equation}
\citep{LL87}, where $v$ is the flow velocity, and $h$ is the enthalpy. The gravitational potential has been neglected, as it is expected to be unimportant at distances of hundreds of AU from the central star. The coupling of acceleration and expansion arises from energy stored as enthalpy being transferred to kinetic energy. However, enthalpy is unimportant in these regions of protostellar outflows \citep[e.g.,][]{Z07}. There may be some exceptions to this assumption, such as in the post-shock cooling region of the recollimation shock where the temperature $\gtrsim 1\textrm{ MK}$ (\S\ref{sec:D-recoll}), but the enthalpy in the region of the flow that is observed to be expanding and accelerating is unimportant. For an isothermal jet, it can be shown via dynamical calculation that the inferred pressure gradient in the DG Tau jet is incapable of accelerating the flow (Appendix \ref{app:turbjet}). Therefore, a purely hydrodynamic acceleration cannot occur in the DG Tau jet. One possible alternative would be the presence of a core, hot flow nested within the jet, formed from or indicated by the presence of the hot stationary X-ray shock in the jet. This material could then accelerate the jet via thermal pressure. The theoretical plausibility of this model is difficult to ascertain, due to the strong dependence of the X-ray material cooling length on both density and shock velocity \citep{GML09}. However, the close proximity of the stationary X-ray and [Fe II] features ($18\textrm{--}30\textrm{ AU}$ separation) suggests the hot shocked material cools over this distance, and would therefore be incapable of driving acceleration at hundreds of AU from the central star.

Magnetic fields sufficiently modify the flow dynamics in a way that could, in principle, provide a mechanism for acceleration to occur (Appendix \ref{app:Baccel}). A tangled magnetic field within the jet may accelerate the jet by the conversion of Poynting flux to kinetic energy. The DG Tau jet is observed to accelerate from $v_0\approx 215\kms$ to $v\approx 315\kms$ over the region $\sim 115\textrm{--}260\textrm{ AU}$ from the central star, and expands from a diameter of $2R_0\approx 20\textrm{ AU}$ to $2R\approx 30\textrm{ AU}$ (Fig.~\ref{fig:bluepower}). For an initial electron number density of $n_{\textrm{e},0}=2\times 10^4\textrm{ cm}^{-3}$, equation (\ref{eq:expaccel}) gives a required initial magnetic field strength $B_0=49\textrm{ mG}$ at a distance of $\sim 115\textrm{ AU}$ from the central star in order to facilitate the coupled acceleration-expansion of the jet. The scaling relationship between density and magnetic field, equation (\ref{eq:Bandrho}), then yields a magnetic field strength of $31\textrm{ mG}$ at the end of the acceleration region, where the electron density has decreased to $\sim 1\times 10^{4}\textrm{ cm}^{-3}$.

A field strength of several tens of mG is plausible, but unlikely, in this region of the outflow. Modelling of the shocks in DG Tau \citep{LFCD00} suggests that the shock velocties in DG Tau are $\lesssim 100\kms$. However, the field strengths calculated above imply an Alfv\'{e}n velocity in the jet of $\sim 315\textrm{--}350\kms$, which is inconsistent with the inferred shock velocities. Even if the tangled field were perfectly isotropic, with an effective Alfv\'{e}n speed of $\sim 105\textrm{--}115\kms$ in any direction, this speed would still be too high to easily allow for shocks of the velocity determined by \citet{LFCD00}. Therefore, we conclude that magnetic acceleration beyond the recollimation shock is unlikely in the DG Tau jet.

In the absence of a source of extra kinetic energy for the jet, we conclude that velocity variations are the most likely cause of the observed ``acceleration'' of the DG Tau jet. We have argued above (\S\ref{sec:D-blueknots}) that these velocity variations are the cause of the moving knots in the DG Tau jet. However, the irregularity in knot ejection intervals and knot proper motions suggests that the underlying jet velocity variation is also irregular. Further time monitoring of DG Tau is necessary to determine the parameters of this variation.

\subsubsection{Rotation}\label{sec:D-bluerot}

The search for jet rotation has been an important component of YSO outflow studies in recent times. The unambiguous determination of rotation in a YSO jet would provide direct evidence that the outflow extracts angular momentum from the circumstellar disc, and offer an answer to the angular momentum problem in star formation \citep[e.g.,][]{B10}. An accurate measurement of the jet rotation would also allow an alternate estimate of the extent of the wind-launching region in the disc \citep[\S\ref{sec:D-launch};][]{Be02,Ae03}.

\begin{figure}
\centering
\includegraphics[width=\columnwidth]{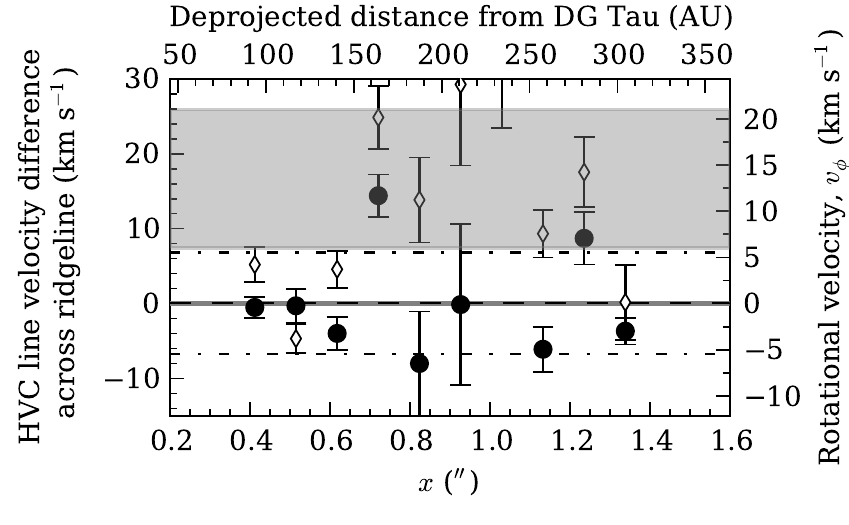}
\caption{Velocity differences across the approaching jet ridgeline. Velocities are taken from the third spaxel above and below the jet ridgeline at each position along the outflow axis, and then subtracted. These differenced velocities are shown as large, filled circles. The average of these differences is shown by the dashed line; the estimated $1\sigma$ uncertainty in this average is denoted by the dot-dashed lines. A velocity difference of $0\kms$ is shown by the thick grey line. Open diamonds show the velocity differences found by the same procedure, but forming differences about the large-scale outflow axis. The grey shading represents the rotational velocities reported by \citet{Ce07}. The rotational velocity is calculated from the velocity difference via the formula $v_\phi=\Delta v/(2\sin i)$, where $i$ is the jet inclination to the line-of-sight \citep{Ce07}.}\label{fig:bluejetrot}
\end{figure}

Our data have been investigated for a rotation signature using a method based on that of \citet{Be02}. If the jet is rotating, the gas on either side of the jet axis will emit lines with slightly different Doppler shifts. At every point along the outflow axis, the fitted HVC line velocities of the third spaxel above and below the jet ridgeline were differenced, covering $0\farcs 11\textrm{--}0\farcs 16 \approx 15\textrm{--}22\textrm{ AU}$ on either side of the jet. The upper limit of this range was chosen to correspond to the greatest observed jet diameter of $\lesssim 40\textrm{ AU}$ (Fig.~\ref{fig:bluepower}(d)). The lower limit of the range was chosen to minimise the beam-smearing of rotational measurements identified in the simulations of \citet{Pe04}, by only including spaxels with central offsets similar to or greater than the PSF ($0\farcs 11$). This procedure measures any velocity asymmetry about the jet ridgeline. The resulting velocity differences are shown in Fig.~\ref{fig:bluejetrot}.

There is no clear indication of rotation in our data of the approaching DG Tau jet. At almost all positions along the jet, the velocity differences are $\lesssim 2\sigma$ from $0 \kms$. Furthermore, the velocity differences across the ridgeline change sign along the jet, which is not consistent with bulk jet rotation. The average velocity difference across the jet ridgeline for all measured locations is $0.0\pm 6.8\kms$, corresponding to a rotational velocity of $v_\phi=0.0\pm 5.5\kms$ after correction for the jet inclination, $i$, by the formula $v_\phi=\Delta v / (2\sin i)$ \citep{Ce07}. This result refutes the lowest rotational velocity claimed by \citet{Ce07} to $0.88\sigma$, and implies an upper limit on observable rotation in the DG Tau jet of $6\kms$. The large uncertainties $\sim 1\arcsec$ from the central star are due to the low signal-to-noise ratio and some spurious line component fits in that region.

\begin{figure*}
\centering
\includegraphics[width=\textwidth]{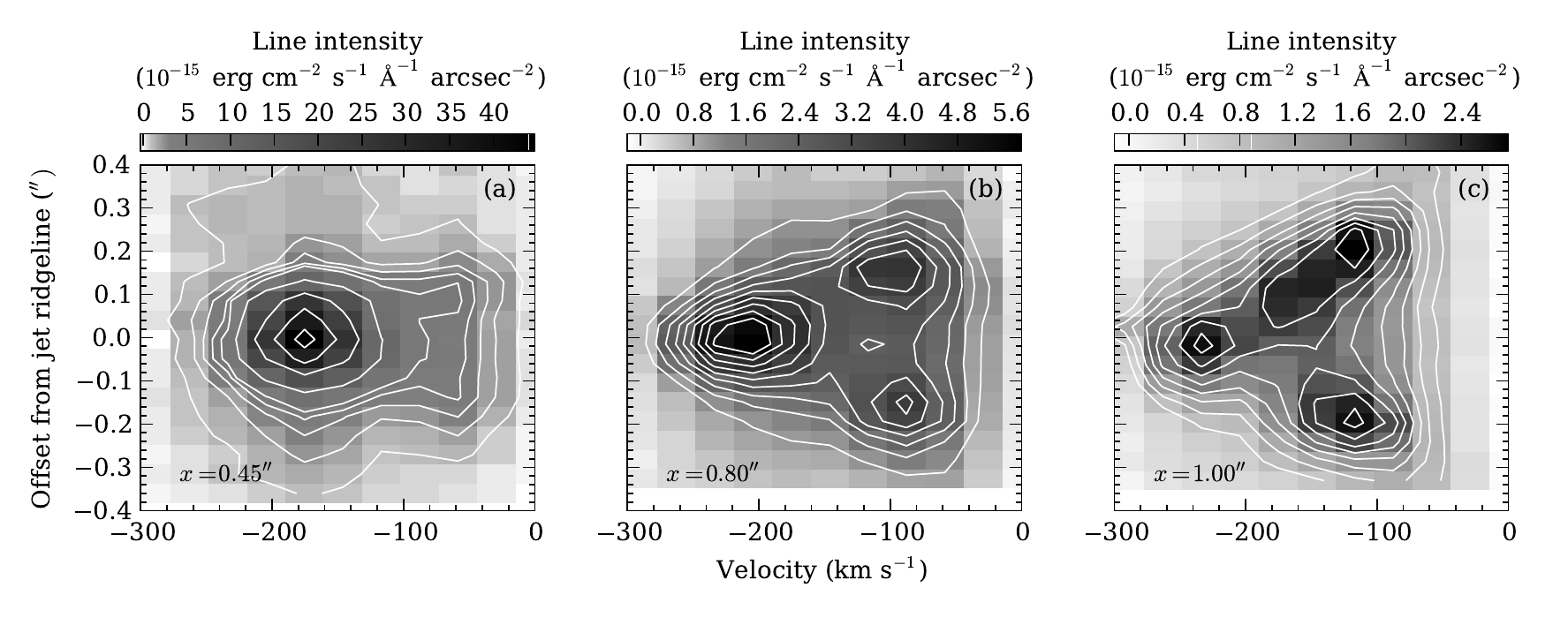}
\caption{Cross-jet position-velocity diagrams of the blueshifted DG Tau outflow at (a) $0\farcs 45$, (b) $0\farcs 80$ and (c) $1\farcs 00$ along the jet. The position offset shown is measured from the position of the jet ridgeline at that distance from the central star. Indicative contours are plotted with the following levels: \\
(a) $[1, 2, \ldots , 5, 10, 20, 30, 40]\times 10^{-15}$ erg cm$^{-2}$ s$^{-1}$ \AA$^{-1}$ arcsec$^{-2}$;\\
(b) $[1.0, 1.5, \ldots , 5.0]\times 10^{-15}$ erg cm$^{-2}$ s$^{-1}$ \AA$^{-1}$ arcsec$^{-2}$;\\
(c) $[0.75, 1.0, \ldots , 2.5]\times 10^{-15}$ erg cm$^{-2}$ s$^{-1}$ \AA$^{-1}$ arcsec$^{-2}$.}\label{fig:bluerotPV}
\end{figure*}

An alternative method for detecting rotation in protostellar outflows is the analysis of position-velocity (PV) diagrams. A rotating jet will show a 'tilted' PV diagram-profile when a spectrograph slit is placed along the cross-jet direction \citep[e.g.,][]{Pe04,Ce04,Ce07}. We have formed cross-outflow PV diagrams of the approaching DG Tau outflow (Fig.~\ref{fig:bluerotPV}) by extracting vertical `slices' of IFU data at positions intermediate between the moving jet knots, $0\farcs 45$, $0\farcs 80$ and $1\farcs 00$ along the jet. We observe that there is no clear, consistent `tilt' in any of these profiles, particularly in the high-velocity component. The intermediate-velocity component may show some small `tilt' at both $0\farcs 8$ and $1\farcs 0$ from the central star, but the direction of this tilt, which corresponds to the inferred direction of rotation, is not the same. Therefore, we conclude that our data do not support the detection of rotation in the approaching DG Tau outflow. We suggest that the any rotation signature originally present in the jet may be degraded by passage through the strong recollimation shock in the jet channel (\S\ref{sec:D-recoll}). 

We now briefly discuss two possible systematic uncertainties in our data. The first is uneven slit illumination, as described by \citet{B02} and \citet{Mae03}. The effect of uneven slit illumination is to create a spurious velocity offset between two positions along the slit due to the convolution of the velocity profile with the pixel width and slit width. However, this is only an issue if the slit width is greater than the PSF width. In our case, our effective slit width for cross-outflow slits ($0\farcs 103$) is comparable to our measured PSF ($0\farcs 11$), so we predict that the impact of this effect on our results will be small. This has recently been confirmed for similar observations of DG Tau in the $K$-band obtained using SINFONI, a similar instrument to NIFS, where it was determined the effect of uneven slit illumination was less than $2\kms$ \citep{A-Ae14}. Second, we must consider the possible effect of residual velocity calibration effects along individual slitlets. This was analysed by \citet{Be08}, and the effect was found to have a magnitude of $\pm \sim 3\kms$. This is less than the $1\sigma$ uncertainty on our determination of the rotation velocity of the DG Tau jet; hence, we determine that this effect is likely negligible on the measured velocity differences.

We proceed to investigate previous claims of rotation in the DG Tau approaching jet. The spectral resolution of STIS, the instrument used to make the previous measurements of claimed rotation, is $\sim 25\kms\textrm{ pix}^{-1}$, with Gaussian fitting typically achieving an effective spectral resolution of one-fifth of the velocity sampling when determining line velocities \citep{Ce07}. The measured velocity differences across the jet in previous rotation studies of DG Tau are factors of a few greater than this uncertainty of $\sim 5\kms$ \citep{Ce07}, which implies that a real velocity asymmetry was detected in previous studies. We shall now investigate possible systematic uncertainties affecting these results.

Our IFU data have an advantage over previous studies in that the location of the jet ridgeline at each downstream position, and the velocity differences at all downstream positions, can be tracked simultaneously. By comparison, the long-slit spectroscopy methods used by \citet{Be02} and \citet{Ce07} can only do one of these, depending on the technique employed. Using multiple slit positions aligned parallel to the large-scale HH 158 outflow axis makes it difficult to locate the centroid of the jet at each downstream position. This requires that the large-scale outflow axis be used as the centre of the jet for forming velocity differences \citep{Be02}. However, it is shown above (\S\ref{sec:jettraj}) that the jet does not follow a linear path along the outflow axis. Repeating our analysis, but forming velocity differences about the large-scale outflow axis, yields an average velocity difference along the jet of $\sim 6\textrm{--}17\kms$, allowing for a $\pm0\farcs 05$ uncertainty in the outflow axis position. We show in Fig.~\ref{fig:bluejetrot} individual velocity differences formed using the large-scale outflow axis as the jet centre (open diamonds in that Figure). These velocity differences are clearly greater than those formed about the jet ridgeline at most downstream positions, especially $0\farcs 7\textrm{--}1\farcs 2$ from the central star.

We note that \citeauthor{Be02} formed velocity differences using the \emph{intermediate}-velocity component of the approaching outflow. We investigate this measurement to demonstrate the importance of our IFU-based method for measuring rotation. First, repeating the analysis described here on the intermediate-velocity component yields the same result as for the jet, with no rotation if the jet ridgeline is taken as the outflow centre, and a rotation velocity of $\sim 5\textrm{--}20\kms$ if the large-scale outflow axis is taken to be the component centre. Second, \citeauthor{Be02} interpreted the IVC as being an intermediate-velocity wind, whereas we have leveraged the capabilities of integral-field spectroscopy to interpret the IVC as a turbulent entrainment layer (\S\ref{sec:D-entrain}). Any rotation signature in such a layer is likely to be masked by the turbulent motion of the entrained gas. We conclude that not centring the velocity difference measurements on the local ridgeline position introduces a possible systematic error in the \citet{Be02} IVC rotation claim.

Conversely, placing the slit across the jet at one downstream position allows for the jet centroid position to be accurately determined \citep{Ce04,Ce07}. However, this measurement provides a velocity difference at only one position along the outflow axis. It can be seen in Fig.~\ref{fig:bluejetrot} that the velocity difference across the jet at any one position may not be an accurate representation of the velocity difference profile of the jet as a whole. Indeed, when the procedure was repeated over multiple epochs for the YSO RW Aurigae, it was found that the cross-jet velocity difference at the sampled position was time-varying on scales of six months \citep{Ce12}, and hence any one measurement of cross-jet velocity difference at one downstream position cannot be reliably used to ascertain the presence of rotation.

The measurement of rotation in YSO jets and outflows is a key piece of evidence supporting MHD disc winds as the driving mechanism, and an important diagnostic in attempting to measure their launch radii \citep{Be02,Ae03,FDC06,Ce04,Ce07}. The non-detection of rotation may be interpreted as weakening the evidence for MHD disc-driven winds. However, we emphasise that other effects may obscure the detection of rotation. Specifically, both the passage of the jet through the recollimation shock (\S\ref{sec:D-recoll}), and the presence of a turbulent entrainment layer (\S\ref{sec:D-entrain}) kinematically process the jet and/or induce turbulence, masking or destroying any rotation that is originally present. In order to definitively confirm or refute jet rotation, it is necessary to either attempt to measure jet rotation upstream of the recollimation shock, investigate rotation in jets without recollimation shocks if they exist, or await higher-resolution integral-field units \citep[e.g.,~GMTIFS;][]{PJMe12} that will allow for the undisturbed jet core to be resolved. 

\subsection{Entrainment Region}\label{sec:D-entrain}

The presence of an intermediate-velocity component (IVC) in the DG Tau blueshifted outflow has been noted by many authors. This component is typically interpreted to be emitted by a less-collimated MHD wind accelerated from the disc around DG Tau, from a radius of a few AU from the central star \citep{Be00,De00,Pe03b,Ae03}. \citet{Pe03b} suggested that at least some part of the DG Tau IVC emission is due to entrainment of such a disc wind by the high-velocity jet, based on the expansion of the IVC as it progresses downstream.

[Fe II] emission is generated by shock interactions \citep{Ne02}. This raises the question as to how a steady, poorly-collimated disc wind radiates in [Fe II]. The HVC radiates predominantly due to the presence of shock-excited knots in the jet (\S\ref{sec:knots}), but no such structures appear in the IVC (Fig.~\ref{fig:bluevelfit}(d)). Sideways ejection of material from the jet knots is also ruled out as the source of shock excitation of the IVC, given the lack of discernible IVC emission enhancements at the knot positions. The formation of a turbulent, shocked entrainment layer between the high-velocity jet and either a wide-angle disc wind, or the ambient medium into which the outflow is emerging, would provide the excitation necessary to dissociate molecules in the wind/ambient medium, and produce [Fe II] emission. We therefore investigate the possibility that the IVC represents a turbulent, shocking entrainment layer.

Entrainment, which is also referred to as turbulent mixing, can occur at two distinct locations within a jet. \emph{Lateral} entrainment occurs along the jet walls, as the fast-moving jet material flowing along the interface pulls the slower-moving/stationary ambient material into a turbulent mixing layer \citep[e.g.,][]{CR91,RCC95}. \emph{Head}, or \emph{prompt}, entrainment is the term used to describe the pushing and mixing that occurs at the head of the jet in a bow shock \citep{RC97}. The head of the approaching DG Tau outflow is at least several arcseconds from the central star \citep{EM98,MR04,MRF07}, so we consider lateral entrainment only. However, full jet flow simulations show that the leading jet bow shock will push aside the ambient medium when the jet is first launched, forming a bubble that keeps the ambient material away from the jet walls \citep{TR95,LRW99}. Therefore, recent models of lateral entrainment apply special conditions to the ambient medium, e.g.~an ambient flow perpendicular to the jet \citep{L-CR10}, to bring the jet and the surrounding medium into contact. There is no evidence for such flows existing in the DG Tau system.

It is often suggested that the high-velocity jets driven by YSOs are nested within a lower-velocity wind \citep[e.g.,][]{Pe03}. Such a wind would come into contact with the jet, and provide a constant supply of molecular material with which to form a mixing layer. This would remove the requirement to apply special conditions to the ambient medium to facilitate entrainment. This scenario was proposed by \citet{Pe03b} as the partial origin of the blueshifted IVC they observed in the DG Tau outflow. Below, we provide evidence that a poorly-collimated molecular disc wind does exist, and argue that the blueshifted IVC is predominantly emitted by a turbulent mixing layer.

\subsubsection{Origin of the Near-Side H$_2$ Region}\label{sec:D-blueH2}

\begin{figure}
\centering
\includegraphics[width=\columnwidth]{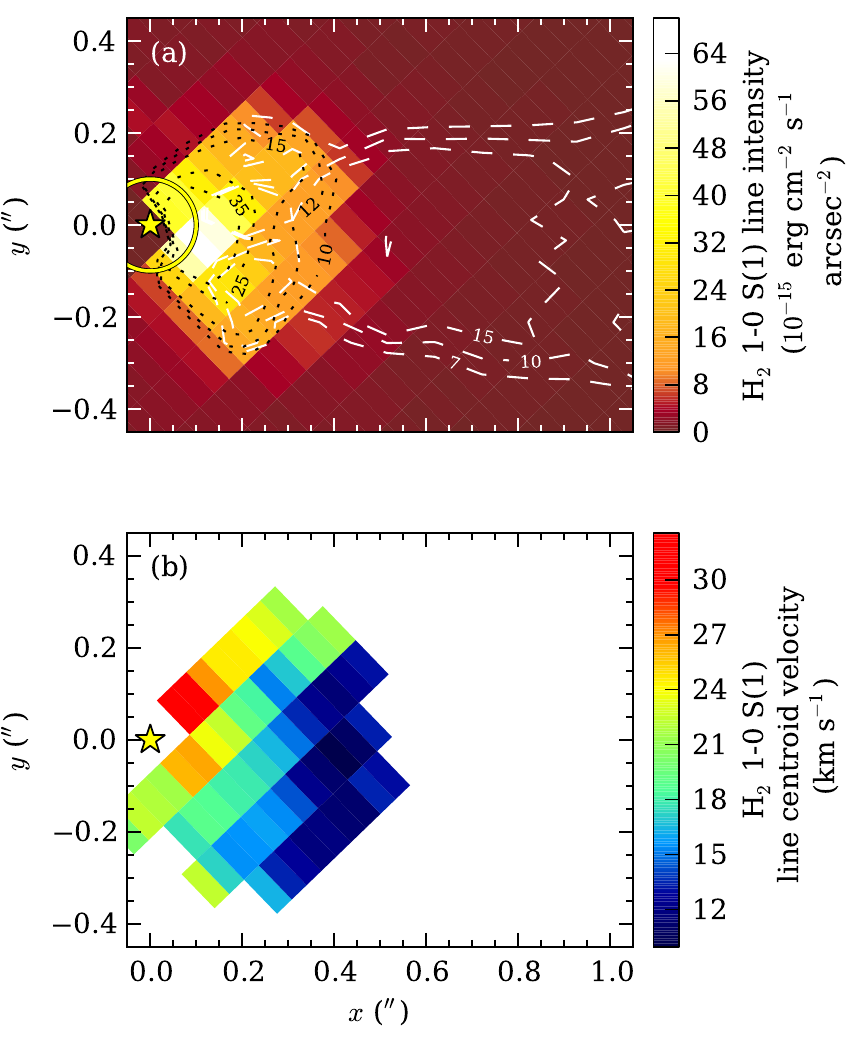}
\caption{H$_2$ 1-0 S(1)  2.1218 $\mu$m emission in the approaching DG Tau outflow. (a) H$_2$ 1-0 S(1) 2.1218 $\mu$m integrated emission flux, formed over the velocity range $-100$ to $60\kms$. Dotted lines (black) show contours of this emission. Overlaid as dashed lines (white) are three contours of fitted [Fe II] 1.644 $\mu$m IVC line intensity (Fig.~\ref{fig:bluevelfit}(d)). Contours are labelled in units of $10^{-15}$ erg cm$^{-2}$ s$^{-1}$ arcsec$^{-2}$. (b) Line velocity centroid of H$_2$ 1-0 S(1) 2.1218 $\mu$m emission in each spaxel, as determined by single-component Gaussian fitting. The velocities quoted are blueshifted velocities, and are adjusted for the stellar velocity, as determined from photospheric absorption line fitting. In both panels, the position of the central star and the position of the occulting disc ((a) only) are shown by a yellow star and circle, respectively.}\label{fig:FeIIH2overlay}
\end{figure}

The extended H$_2$ 1-0 S(1) 2.1218 $\mu$m line emission from the near side of the DG Tau circumstellar disc takes on a bowl-shaped morphology, as shown in Fig.~\ref{fig:FeIIH2overlay} \citep[also,][]{Be08}. This H$_2$ emission was interpreted by \citet{Te04} as being from a warm, wide-angle molecular wind encasing the inner regions of the HH 158 outflow. Data on the approaching H$_2$ emission obtained in the ultraviolet by \citet{Ae02a} and \citet{He06} are consistent with this explanation, and \citet{Be08} and \citet{A-Ae14} also concluded that their data support this assertion. We provide further evidence below that this emission comes from a wider-angle molecular wind.

To investigate the velocity structure of the H$_2$ emission, spectral Gaussian fits were made to the H$_2$ 1-0 S(1) 2.1218 $\mu$m line at every position in the $K$-band data cube, using the same method applied to the [Fe II] 1.644 $\mu$m line in the $H$-band data cube (\S\ref{sec:fitted}). Fits were restricted to a single line component. Furthermore, a lower signal-to-noise ratio threshold of 2.5 was applied to fits in the $K$-band data cube. The line velocities were adjusted to account for the systemic stellar velocity, based on absorption line fits to the Na I and Ca I doublets visible in the $K$-band stellar spectrum (Fig.~\ref{fig:stellar05}(b)). The resulting line centroid velocity profile is shown in Fig.~\ref{fig:FeIIH2overlay}(b).

The near-side H$_2$ emission is all blueshifted with respect to the systemic velocity (Fig.~\ref{fig:FeIIH2overlay}). This eliminates the circumstellar disc surface as the origin of the emission, through either emission or scattering by the disc surface. If the emission was produced or scattered by the disc surface, it would be expected to have zero line velocity with respect to the systemic velocity, with a small asymmetry of $\sim$ a few$\kms$ about the outflow axis, caused by the rotation of the disc. Such an asymmetry is present $\sim 0\farcs 2$ along the outflow axis, but it is too large, $\sim 9\kms$, to represent disc rotation. It also shows the opposite rotational sense to the known rotation direction of the DG Tau circumstellar disc \citep{Tes02}. We determine line centroid velocities between $-10$ and $-30\kms$ for the H$_2$ emission, which are larger than the $-2.4\pm 18\kms$ reported by \citep{Be08}. This discrepancy results from \citeauthor{Be08} reporting the line centroid velocity of all the emission, which includes the $\sim 0\kms$ H$_2$ emission on the far side of the disc. Indeed, our approaching centroid line velocity determinations are mostly within the uncertainties given by \citet{Be08}. Our measurements may also suffer from the effects of uneven slit illumination, as described by \citet{A-Ae14}. Indeed, those authors report lower blueshifted velocities ($\sim 5\kms$) for the majority of the H$_2$ emission.

The H$_2$ 1-0 S(1) line velocity map provides clues as to the nature of this outflow. 
The line velocity peaks near the central star, and decreases with distance along the outflow axis. This effect was also observed by \citet{A-Ae14}.
We interpret this to be the profile of a poorly-collimated wind.
The higher approaching line velocities near the base of the wind correspond to where the wind has just been launched, and has yet to be collimated into the outflow direction. 
The gas on the near side of the wind is therefore flowing towards the observer, increasing the line-of-sight velocity component. 
As the flow becomes collimated, the gas flows in the outflow direction, and hence the line-of-sight velocity component becomes smaller. 

\begin{figure}
\centering
\includegraphics[width=\columnwidth]{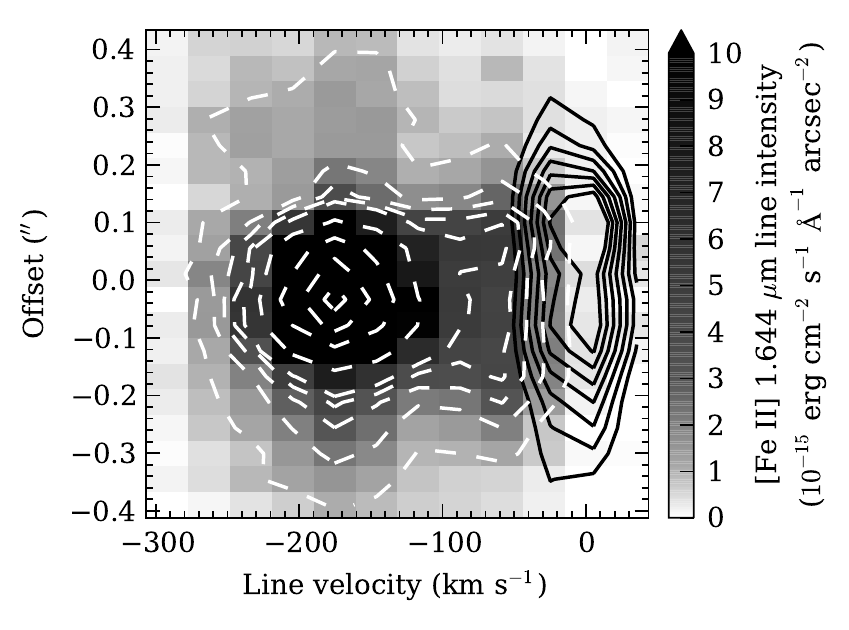}
\caption{Cross-outflow position-velocity diagram of H$_2$ 1-0 S(1) 2.1218$\mu$m and [Fe II] 1.644 $\mu$m emission in the approaching DG Tau outflow, formed $0\farcs 425$ from the central star. The [Fe II] 1.644 $\mu$m emission is shown in greyscale. [Fe II] 1.644 $\mu$m emission contours (white, dashed) are placed at levels of $[1,2,\ldots,5,10,20,30,40]\times 10^{-15}\textrm{ erg cm}^{-2}\textrm{ s}^{-1}\textrm{ \AA}^{-1}\textrm{ arcsec}^{-2}$. H$_2$ contours (black, solid) are placed at levels of $[0.4,0.6,\ldots,1.6]\times 10^{-15}\textrm{ erg cm}^{-2}\textrm{ s}^{-1}\textrm{ \AA}^{-1}\textrm{ arcsec}^{-2}$. The $K$-band data have been re-gridded onto the $H$-band pixel grid by linear interpolation.}\label{fig:FeIIH2PV}
\end{figure}

We search for a kinematic link between the H$_2$-emitting material and the IVC of the [Fe II] emission. Fig.~\ref{fig:FeIIH2PV} shows a position-velocity diagram of both the [Fe II] and H$_2$ emission at the observable edge of the latter. This diagram tentatively suggests that the `wings' of the [Fe II] IVC form a `bridge' between the H$_2$ emission, and the higher-velocity [Fe II] emission, which may be indicative of shearing and entrainment. We also note that the [O I] LVC reported by \citet{Ce07} may further spatially and kinematically link the H$_2$ and [Fe II] HVC emitting material \citep{A-Ae14}. This suggests that all three components are kinematically linked, supporting our interpretation of the [Fe II] IVC as an entrainment layer between the molecular wind and the high-velocity jet.

\subsubsection{Requirements for Lateral Entrainment}\label{sec:D-entrainreq}

Lateral entrainment occurs via instabilities that form along the walls of the jet and cause turbulent mixing of jet and ambient material. The relevant instability in the formation of mixing layers is the Kelvin-Helmholtz (KH) instability \citep{C61}. Velocity shears are well-known to be stabilised against the KH instability for a Mach number difference, $M_\Delta$, between the flows of $M_\Delta \gg 1$ \citep{T08}. Hydrodynamic simulations have shown that entrainment is unimportant in jet flows with a Mach number difference of $M_\Delta >6$ \citep{Ce94}. It has been shown analytically that hydrodynamic shear layers are stabilised against the KH instability if $M_\Delta\geq\sqrt{8}\approx 2.8$ for disturbances propagating in the jet flow direction \citep{T08}. More generally, taking into account instability modes which propagate at an angle $\phi$ to the outflow, the criterion for stability is $M\cos\phi < \sqrt{8}$ \citep{FM63}. This means that some KH instability modes may be unstable for $M_\Delta>\sqrt{8}$, permitting more modest entrainment at higher Mach number differences. However, given that the DG Tau jet is highly supersonic, with $M_\Delta\approx M_\textrm{jet}\sim 18\textrm{--}27$ for a monatomic jet at temperature $T=10^4\textrm{ K}$, lateral entrainment is unimportant if the DG Tau jet is purely hydrodynamic.\footnote{$M_\Delta=M_\textrm{jet}$ if the ambient material is at rest with respect to the star-disc system. If the ambient material is the less-collimated molecular wind, it is significantly slower than the jet, such that $M_\Delta\approx M_\textrm{jet}$.}

Magnetic fields can permit lateral entrainment to occur in highly supersonic jets. The effects of magnetic fields on the KH instability in shear layers are complex, and are sensitive to both the initial physical conditions of the flow, and the orientation of the magnetic field lines with respect to the flow and shear directions \citep{C61,T08}. Consider a slab shear layer between magnetised compressible gases in the $y,z$-plane, with the velocity shear occurring in the $y$-direction, and the fast-moving gas on one side of the shear layer flowing in the $z$-direction. The shear layer then extends infinitely in the $x$-direction. There are three basic magnetic field orientations that illustrate the complexities at hand. First, if the magnetic field is parallel to both the shear interface and the flow direction, that is, $\vec{B}=B\hat{\vec{z}}$, then the shear layer is stabilised against the KH instability if $v_\textrm{A} \geq c_\textrm{s}$ \citep{C61,RE83}. This condition is satisfied for the magnetic field strengths inferred for protostellar jets \citep{LFCD00,He07}. Second, a magnetic field perpendicular to both the interface and flow direction, that is, $\vec{B}=B\hat{\vec{y}}$, has no effect on the suppression of the KH instability \citep{C61}.

Consider an astrophysical jet described in cylindrical coordinates $(r,\phi,z)$, flowing in the $z$-direction. In this case, the shear layer between the jet and the ambient medium (or an encasing wind) will be in the $(\phi,z)$-plane. Beyond the Alfv\'{e}n surface a few to tens of AU above the circumstellar disc, the magnetic field in the outflow will be predominantly toroidal \citep{He07,Z07}. Then, the most physically accurate two-dimensional shear layer approximation for a protostellar jet is that with a magnetic field parallel to the shear interface, but perpendicular to the flow direction, such that $\vec{B}=B\hat{\vec{x}}$. For this field configuration, the KH instability criterion is as for a purely hydrodynamic jet, but with the Mach number difference across the shear layer determined with respect to the quadrature sum of the sound and Alfv\'{e}n speeds, $\sqrt{c_\textrm{s}^2+v_\textrm{A}^2}\approx v_\textrm{A}$ for $v_\textrm{A} \gg c_\textrm{s}$ \citep{MP82,RE83}. The Alfv\'{e}nic speed becomes the effective sound speed.

To destabilise the interface between the jet and ambient wind in DG Tau, the magnetic field encompassing the jet would need to result in an Alfv\'{e}n jet velocity of $75\textrm{--}115 \kms$. Such an Alfv\'{e}n velocity is low enough to allow for the formation of shocks with the velocities inferred by \citet{LFCD00}. The upper limit on the required magnetic field strength is $7.5\textrm{--}11\textrm{ mG}$, based on our determination of the density of the DG Tau jet (\S\ref{sec:D-bluepower}). This field strength is an order of magnitude greater than that inferred by \citet{LFCD00} for DG Tau from shock modelling, but is an order of magnitude less than the magnetic field strength in protostellar jets considered reasonable by \citet{He07}.\footnote{Incidentally, this magnetic field is also significantly weaker than the field strength necessary to cause extended acceleration in the jet (\S\ref{sec:D-blueaccel}).} As stated above, shear layer disturbances propagating at an angle to the flow direction will not be stabilised until higher Mach number differences are reached \citep{FM63}. A weaker field could facilitate a lower entrainment rate in the DG Tau jet. Indeed, the DG Tau jet does not become fully turbulent over the region where entrainment is occurring (\S\ref{sec:blueoutflowHVC}), suggesting that only moderate turbulent mixing is occurring. Therefore, we consider the magnetic field strength necessary to enable turbulent entrainment to be physically reasonable, and conclude that the magnetic field providing collimation to the DG Tau jet also allows the jet to entrain material from the ambient wind.

\subsubsection{Relationship to Large-Scale Molecular Outflows}

One of the most striking features of Class 0 and Class I protostars are large-scale bipolar molecular outflows detected in millimetre rotational transitions of CO \citep{S94b}. Such outflows were first detected around the protostar L1551 IRS 5 \citep{SLP80}, and were quickly identified as being common in star-forming regions \citep{RB97}. The masses of these outflows are greater than the mass of the driving protostar, implying that the outflow must be composed of swept-up material \citep{MC92}. Typically these outflows have ages $\sim 5\times 10^{3\textrm{--}4}\textrm{ yr}$ \citep{MC93}, and the long cooling time of the CO molecule provides a history of the outflow \citep{R00}. These swept-up shells are generally interpreted as being driven by prompt entrainment from an outflow bow shock \citep{CRG97,De97,RB97,AG02,Se06}.

We argued above for the presence of lateral entrainment in the DG Tau microjet. Such entrainment provides another candidate source for the momentum in the large-scale swept-up molecular outflows. Previous studies argued against lateral entrainment as a driving mechanism for CO outflows \citep[e.g.,][]{RC93,De97,RB97}. These studies relied on the argument that the KH instability would not develop in protostellar jets; however, we argued above that in fact, this is possible when magnetic effects are taken into account (\S\ref{sec:D-entrainreq}). Lateral entrainment would be particularly useful in objects such as HH 286, where the molecular outflow ends closer to the protostar than the location of the first optical Herbig-Haro object, indicating the jet has pushed past the head of the CO outflow. Hence the jet can no longer drive the CO outflow in a snowplow fashion \citep{Se06}, and lateral entrainment becomes a possible CO outflow driving mechanism. However, it should be noted that in many recent high-angular resolution observations of molecular outflows, the structure and kinematics of the outflow has favoured the bow-shock driving model \citep{GG99,Leee02}, and the driving in such an object may be from a wide-angle wind instead of a well-collimated jet \citep[][and references therein]{Are07}. However, lateral entrainment may still provide some contribution, albeit small, to the driving of CO outflows.

There is no detected CO outflow associated with DG Tau. However, DG Tau is currently transitioning between evolutionary Class I and Class II \citep[][]{Pe03b,WH04}, and any CO outflow that was previously present must have cooled to the point where it is no longer emitting. A decrease in $^{13}\textrm{CO}$ column density $\sim 4000\textrm{ AU}$ from the central star indicates that a major part of the disc-shaped envelope around DG Tau has already been blown away, and the molecular outflow responsible for the dispersion is no longer visible \citep{KKS96a}. Attempting to locate lateral entrainment in the microjets of younger YSOs that drive CO outflows would be difficult, due to the significant extinction towards these highly embedded objects. Therefore, numerical simulations will be useful to test the viability of lateral entrainment as a mechanism for driving CO outflows. Such models would need to account for the magnetic fields in and around the outflows from the YSO in order to facilitate lateral entrainment.

\section{Conclusions}\label{sec:concl}

We have investigated the YSO DG Tauri, and its associated outflows, in detail using $H$- and $K$-band data from the NIFS instrument at Gemini North taken on 2005 Oct and Nov. The $H$-band stellar spectrum shows significant photospheric absorption features, in contrast to previous studies of DG Tau that showed a veiled continuum spectrum. The $K$-band stellar spectrum also shows significant photospheric absorption features, as well as CO $\Delta v=2$ bandheads in absorption. These bandheads appear to oscillate between absence, emission and absorption, depending upon the observing epoch. The lack of a veiling continuum, and the absence of CO bandheads in emission, suggests that DG Tau was in a low accretion rate phase during this observation epoch. This is consistent with our observation epoch being between periodic outflow episodes.

Two regions of extended emission were detected about the central star, on opposing sides of the circumstellar disc. Three distinct emission components were observed in the blueshifted, or approaching, outflow, out to a distance of $1\farcs 5$ from the central star:
\begin{description}
\item[{\bf High-velocity jet.}] A high-velocity, well-collimated central jet is seen as the high-velocity component (HVC) of [Fe II] 1.644 $\mu$m line emission. A stationary emission knot is observed at the base of the outflow, $\sim 0\farcs 2$ from the central star. We interpret this feature as a jet recollimation shock, based on comparison with X-ray \citep{Ge05,Ge08,Ge11,SS08,GML09} and FUV \citep{Sce13} observations. The entire jet shocks to a temperature of $\sim 10^{6}\textrm{ K}$, but only a small region of this shock emits strongly in X-rays \citep{Be11}. The jet material then cools as it flows downstream. Using the pre-shock flow velocity inferred from X-ray observations of $\sim 400\textrm{--}700\kms$, we calculate that the innermost streamlines of the jet are launched from a radius of $0.01\textrm{--}0.15\textrm{ AU}$ from the central star, assuming an MHD disc wind. This range of launch radii could correspond to either a disc wind or an X-wind. The post-recollimation-shock jet is seen as the HVC of [Fe II] emission, having been decelerated to $\lesssim 215\kms$. The jet follows a non-linear path in the NIFS field, and changes in both velocity and diameter along its length. After accounting for the wandering jet trajectory, we find no evidence of rotation in the jet, which is consistent with the effects of passage through a strong recollimation shock.

Two moving jet knots are detected, and labelled knots B and C. Knot B is seen to move at $0\farcs 17\pm 0\farcs 01\textrm{ yr}^{-1}$, much slower than previously observed knots in the DG Tau jet. Knot C is only observed in our 2005 epoch data, and hence we are unable to reliably constrain the proper motion and launch date of that feature. Our data suggest that the interval between knot ejections is non-periodic, and the velocity of the ejected knot varies between ejection events. The jet velocity increases from $215\kms$ to $315\kms$ deprojected between the moving knots, which after the elimination of alternative explanations we interpret to be the result of intrinsic jet velocity variations. These velocity variations are likely the cause of the formation of the moving knots.

\item[{\bf Entrainment region.}] A second outflow component in [Fe II] 1.644 $\mu$m emission was separated from the jet emission, using a multi-component Gaussian line fitting routine based on the statistical $F$-test. This intermediate-velocity component (IVC) takes the appearance of a wider-angle flow. Comparison to the molecular wind detected in the $K$-band (see below), as well as consideration of the excitation method of the forbidden [Fe II] lines, suggests that this component represents a shocking, turbulent entrainment layer between the central jet and the wide-angle molecular wind. A magnetic field of with a strength of $\lesssim$ a few mG allows for entrainment to occur by destabilising the jet-wind interface, although careful analysis of the effects of field orientation is required. The presence of lateral entrainment in a YSO outflow provides an interesting alternative driving mechanism for large-scale CO outflows in younger-type YSOs. An analytical model of this entrainment will be presented in a future paper (White et al.~2014c, in preparation).

\item[{\bf Molecular outflow.}] Wide-angle H$_2$ 1-0 S(1) 2.1218 $\mu$m emission was observed on the near side of the DG Tau circumstellar disc, as reported by \citet{Be08}. Line velocity mapping of this emission indicates that it is most likely due to a wide-angle molecular wind, which agrees with the conclusions of \citeauthor{Be08} and \citet{A-Ae14}.
\end{description}

A receding outflow was detected on the far side of the DG Tau circumstellar disc. This disc obscures our view of this outflow out to $\sim 0\farcs 7$ from the central star, corresponding to an outer disc radius of $\sim 160\textrm{ AU}$. The redshifted outflow takes the form of a bubble-like structure in [Fe II] 1.644 $\mu$m line emission. There is tentative evidence for the presence of an underlying jet, although this cannot be confirmed without further data from later epochs. We will discuss the nature of this structure in a future paper \citep{MCW13b}.

Many of the above conclusions depend on time-varying mechanisms. Further multi-epoch data are therefore required in order to validate these findings. In particular, confirmation of the knot launch period and proper motions requires multi-epoch data taken in the same fashion. It is also of interest to see how the velocity differences across the jet evolve with time, and if any trend attributable to rotation can be identified. Multi-epoch data will also help to settle the question of whether the mass flux and kinetic power of the approaching jet are constant or time-varying. In the future, the advent of 30 m-class telescopes such as GMT will allow for a finer cross-jet sampling, which is necessary to detect complex velocity structures within the jet.

\section*{Acknowledgements}

Based on observations obtained at the Gemini Observatory, which is operated by the 
Association of Universities for Research in Astronomy, Inc., under a cooperative agreement 
with the NSF on behalf of the Gemini partnership: the National Science Foundation 
(United States), the National Research Council (Canada), CONICYT (Chile), the Australian 
Research Council (Australia), Minist\'{e}rio da Ci\^{e}ncia, Tecnologia e Inova\c{c}\~{a}o 
(Brazil) and Ministerio de Ciencia, Tecnolog\'{i}a e Innovaci\'{o}n Productiva (Argentina).

We are extremely grateful for the support of the NIFS teams at
the Australian National University, Auspace, and Gemini Observatory for their tireless efforts during the instrument integration, commissioning and system verification: Jan Van Harmleen, Peter Young, Mark Jarnyk (deceased), Nick Porecki, Richard Gronke, Robert Boss, Brian Walls, Gelys Trancho, Inseok Song, Chris Carter, Peter Groskowski, Tatiana Paz, John White, and James Patao.

We thank the anonymous referee for their detailed comments. M.~White acknowledges the generous travel support from Academia Sinica to attend the conference Star Formation Through Spectroimaging at High Angular Resolution in July 2011, which provided useful information for this study. This work was supported by the Australian Research Council through Discovery Project Grant DP120101792 (R.~Salmeron).

\bibliographystyle{mn2e}
\bibliography{library}

\appendix

\section{The $F$-Test}\label{app:Ftest}

Formally, the $F$-test combines two different methods of computing a $\chi^2$ statistic, and compares the results to determine if their relationship is reasonable. If two statistics following the $\chi^2$ distribution have been determined, then the ratio of the reduced-$\chi^2$ of those distributions is distributed according to the $F$-distribution \citep{BR92}.

Given the additive nature of functions obeying $\chi^2$ statistics, a new $\chi^2$ statistic may be formed by taking the difference of two $\chi^2$ statistics. In particular, consider fitting a model with $m_1$ free parameters (the simpler model) to $N$ data points. Then, the corresponding chi-square value associated with the deviations about the regression $\chi^2_1$ has $N-m_1$ degrees of freedom. Adding another term to the model, with an extra $\Delta m$ free parameters such that $m_2=m_1+\Delta m$, will lead to a corresponding regression $\chi^2_{2}$ with $N-m_2$ degrees of freedom (the more complex model). Forming the ratio of the difference in chi-square values to the more complex model reduced chi-squared forms a statistic that obeys the $F$-distribution,
\begin{equation}\label{eq:Fratio}
F = \frac{(\chi^2_1-\chi^2_2)/\Delta m}{\chi^2_2/(N-m_2)}=\frac{\Delta\chi^2/\Delta m}{\chi^2_2/(N-m_2)}\textrm{.}
\end{equation}
This ratio is a measure of how much the additional term has improved the value of the reduced chi-squared, and should be small if the more complex model does not produce a fit significantly better than the simpler model \citep{BR92}. The $F$-test determines if the improvement in $\chi^2$ between the models warrants the loss of degrees of freedom. If the above ratio equation (\ref{eq:Fratio}) is $\sim 1$, then the change in $\chi^2$ is not significant when compared to the reduction in degrees of freedom, and the more complex model is therefore not a statistically significant improvement, and would be rejected as unjustified.

If the $F$-ratio is significantly greater than one, there are two possibilities. One is that the more complex model is a statistically better fit to the data. However, it is also possible that, by coincidence, noise in the data has taken the form of an extra term to be fitted by the model. To estimate the probability of this, the $F$-distribution is used:
\begin{align}
P_F(F,\nu_1, \nu_2) &= \int_F^\infty P_f(f,\nu_1,\nu_2)\textrm{ d}f\textrm{, where} \\
P_f(f,\nu_1, \nu_2) &= \frac{\Gamma(\nu_1+\nu_2/2)}{\Gamma(\nu_1/2)\Gamma(\nu_2/2)}\left(\frac{\nu_1}{\nu_2}\right)^{\nu_1/2} \nonumber\\
                    &\phantom{=}\times\frac{f^{1/(2(\nu_1-2))}}{(1+f\nu_1/\nu_2)^{1/(2(\nu_1+\nu_2))}}\textrm{,}
\end{align}
where $\nu_1=\Delta m$ and $\nu_2=N-m_2$ are known as the degrees of freedom in the numerator, and degrees of freedom in the denominator, respectively, of equation (\ref{eq:Fratio}). These values characterise the $F$-distribution that has been generated \citep{BR92,Wee07}. This is then a test of whether the coefficient of the extra term in the more complex model is zero. In this formulation, if the probability $P_F(F,\nu_1,\nu_2)$ exceeds some test value (typically 5\%), then one may be fairly confident that the coefficient of the extra term is not zero, and hence the more complex model is a statistically significantly better fit to the data. Otherwise, it is rejected, and the simpler model is retained \citep{BR92}. 

\subsection{Applicability}\label{app:Ftest-applic}

There are two necessary conditions for the proper use of $F$-statistics. The first is that the two models being compared must be nested. The simpler model must be the more complex model with some parameters set to special null values, typically one or zero. This is clearly satisfied when testing for the presence of extra spectral line components, as one may remove the extra line component from the more complex model by setting the component amplitude to zero. The other, less well-known condition, is that the null values of the additional parameters may not be on the boundary of the set of possible parameter values. This is violated when testing for extra emission line components, as the amplitude of the line may not be negative, and hence the boundary of the allowable values for the line amplitude is zero. This is the same as the null value. Hence, the test is being used outside of the formal mathematical definition, and so the underlying reference distribution of the statistic is unknown. One suggested alternative test is Bayesian model checking. However, this requires extensive Monte Carlo simulations to generate the test statistic \citep{Pre02}, which is not practicable for a large quantity of spectra.

The $F$-test will not necessarily produce incorrect results if used to detect extra spectral line components. \citet{Pre02} determined that model-checking with the $F$-test produces a false-positive rate of between 1.5\% and 31.5\%. Furthermore, as an example, they re-analysed the detection of the Fe K line in a gamma-ray burst X-ray afterglow, GRB 970508, which had previously been claimed by \citet{Piro99} based on an $F$-test. Re-analysis of the line detection with Bayesian statistics did not disprove the \citeauthor{Piro99} detection, but confirmed it with a higher significance. \citeauthor{Pre02} also pointed out that the more sophisticated Bayesian methods have their own inherent flaws. Ultimately, there is no `correct' test for all nested model situations; rather, a test appropriate to the particular model and context must be selected \citep{Pre02}.

\section{Dynamical Calculations of a Turbulent Jet}\label{app:turbjet}

We consider a dynamical model for the DG Tau jet, which we summarise here. The jet passes through a recollimation shock $\lesssim 50\textrm{ AU}$ along the outflow channel. This produces a hot X-ray knot \citep{Ge05,Ge08,Ge11,SS08,GML09} for the innermost streamlines, although a large fraction of the surrounding jet gas is also heated to $\sim 10^6\textrm{ K}$ (\S\ref{sec:D-recoll}). The jet rapidly cools to a few $\times 10^{4}\textrm{ K}$ \citep{B02,Me14}, hence we see the jet mainly as an optical/infrared source. The supersonic jet interacts with the surroundings, becoming turbulent and entraining ambient gas (\S\ref{sec:D-entrain}). Since the jet is supersonic, the amount of entrainment and related deceleration is modest. The turbulence associated with the entrainment produces the $50\textrm{--}100\kms$ shocks observed in the DG Tau outflow \citep{LFCD00}. This turbulence also counteracts the radiative cooling of the jet, so the jet gas remains approximately isothermal. This, combined with the relatively flat density gradient within the jet, means the pressure gradient is also modest along the jet, and unable to cause acceleration (\S\ref{sec:D-blueaccel}). 

We now proceed to outline the calculations which support the above description. We adopt an ionisation fraction, $\chi_e$, of 0.3 for the jet \citep{B02,Me14}. We take a helium number density, $n_\textrm{He}=X(\textrm{He})n_\textrm{H}=X(\textrm{He})n_e\chi_e^{-1}$, where $X(He)\approx 0.085$ is the solar helium abundance with respect to hydrogen. Then, the total number density, $n=( 1 + [ 1 + X(\textrm{He})]\chi_e^{-1})n_e$, and the mass density, $\rho=(1+4X(\textrm{He}))\chi_e^{-1}n_em$, where $m$ is the atomic mass unit. In the following, $k$ is the Boltzmann constant, $T$ is the gas temperature, and $\Lambda(T)$ represents the cooling function.

\subsection{Cooling After the Recollimation Shock}

The existence of the recollimation shock means that the jet comes into pressure equilibrium with the surroundings, so a model of a pressure-confined jet is feasible. The jet cools fairly rapidly following this shock, with a cooling length given by,
\begin{align}
l_c &= \frac {3}{2} \, \frac {n}{n_e n_H} \frac {kT}{\Lambda(T)} \, v_{\rm jet} \nonumber\\
    &\approx 38\textrm{ }n_{e,6}^{-1}T_{6.5}\Lambda_{-22.5} \left(\frac{v_\textrm{jet}}{200\kms}\right)\textrm{ AU,}
\end{align}
where subscript numbers denote the quantity in exponential units of that power. When the temperature drops to $\sim 10^5\textrm{ K}$ the cooling becomes even more rapid, so it is not surprising that the jet is seen at optical and infrared wavelengths with a temperature of a few $\times 10^4\textrm{ K}$.

\subsection{Jet Turbulent Velocity}

We now estimate the turbulent velocity within the jet. We take a cylindrical coordinate system, ($r,\phi,z$), with $z$ along the jet axis. Let $\bar \rho$, $\bar p$, $\tilde v_r$ and $\tilde v_z$, be the mean density, pressure and radial and axial velocity components along the jet direction ($z$) and let $\phi_g$ be the gravitational potential. For a jet subject to hydrodynamic turbulence, the $z$-momentum equation for the mean flow is
\begin{equation}
\pd {(\bar \rho \tilde v_z^2)}{z} + \frac {1}{r} \pd{(r \bar \rho \tilde v_r \tilde v_z)}{r} = - \pd {\bar p}{z} 
- \bar \rho \pd{\phi_g}{z} 
-\frac{1}{r} \pd{(r \langle \rho v_r^\prime v_z^\prime \rangle)}{z} 
\end{equation}
where $- \langle \rho v_r^\prime v_z^\prime \rangle$ is the Reynolds stress \citep[see][]{Bi84,KB04}, and angle brackets denote mass-weighted time-averaged quantities according to the \citet{F69} prescription. Primes are used to denote locally fluctuating quantities; bars and tildes denote time-averaged quantities.

For a jet in pressure equilibrium, $p(r,z) = p_{\rm ext}(z)$, the external pressure. For a stellar mass of $0.67 M_\odot$ \citep{HEG95}, the gravitational field is unimportant in the accelerating region. Let $R(z)$ be the jet radius. Then, for a jet which is spreading due to turbulence,
\begin{align}
\langle \rho v_r^\prime v_z^\prime \rangle & \approx \bar \rho \tilde v_z^2 \frac {\dd R}{\dd z} \\
\Rightarrow v^\prime & \approx 110\kms \> \left( \frac{\tilde v_z}{200\kms} \right) \, \left(\frac {\dd R/\dd z}{0.1} \right)^{0.5} \label{v_turb}
\end{align}
The observed value of $\dd R/\dd z \sim 0.05 - 0.1$ so that equation~(\ref{v_turb}) agrees well with the turbulent velocity implied by both the HVC line widths (Fig.~\ref{fig:bluevelfit}(c)) and the results of emission line modelling \citep{LFCD00}.

\subsection{Turbulent Dissipation of Energy}

The rate of production of turbulent energy per unit volume is given by:
\begin{align}
\dot \epsilon_t & = \langle \rho v_r^\prime v_z^\prime \rangle \, \tilde v_{z,r} \\
 & \approx \frac {\bar \rho \tilde v_z^3}{R} \> \frac {\dd R}{\dd z}\textrm{.}
\end{align}
This energy is dissipated and heats the plasma. For DG Tau, the amount of energy produced is of order $10^{-13}\textrm{--}10^{-12}\textrm{ erg s}^{-1}\textrm{ cm}^{-3}$. By comparison, the rate of cooling in the jet, based on a nominal cooling function of $\Lambda(T)=10^{-22}\textrm{ erg cm}^{3}\textrm{ s}^{-1}$ as appropriate for a $\sim 10^{4}\textrm{ K}$ plasma in collisional ionisation equilibrium, is of order $10^{-14}\textrm{ erg s}^{-1}\textrm{ cm}^{-3}$. The estimated heating exceeds the cooling rate, maintaining the jet temperature at $\sim 10^{4}\textrm{ K}$.

\subsection{Pressure-Driven Jet Acceleration}

We aim to determine if the acceleration of the DG Tau jet over the region $0\farcs 5\textrm{--}1\farcs 15$ from the central star could be consistent with the inferred pressure gradient in the jet. There are two possible approaches. The first approach considers the momentum budget in the jet, while the second is based on a Bernoulli equation-type analysis. Both methods show that the pressure gradient in the DG Tau jet is incapable of providing acceleration.

\subsubsection{Momentum Budget}

Let us assume that the jet is in a steady state, and the observed increase in velocity in the jet is the result of acceleration by the pressure gradient. Integrating the momentum equation over the jet cross-section, neglecting the gravitational force, yields
\begin{equation}
\label{diff_momentum}
\frac{\dd}{\dd z} \left[ 2 \pi \int_0^\infty \bar \rho \tilde v_z^2 \> r \> dr \right] = - \frac{\dd\bar{p}}{\dd z} \times A(z)\textrm{,}
\end{equation}
where $A(z)$ is the jet cross-sectional area. This equation integrates to:
\begin{equation}
\label{momentum}
\rho_2 v_2^2 A_2 - \rho_1 v_1^2 A_1 \approx - \int_{z_1}^{z_2} \frac{\dd\bar p}{\dd z} A(z) \> dz\textrm{.}
\end{equation}
All quantities in this equation can be estimated from our observational data. The pressure may be estimated as $p=nkT$. For the DG Tau jet, over the region of increasing jet velocity, the difference in momentum on the left-hand side of equation \ref{momentum}, $1.1\times 10^{25}\textrm{ g cm s}^{-1}$, is two magnitudes of order higher than the average inferred pressure gradient along the jet multiplied by the average jet radius, $2.7\times 10^{23}\textrm{ g cm s}^{-1}$. Therefore, the pressure gradient cannot drive the observed momentum increase of the jet.

\subsubsection{Bernoulli Equation-Type Analysis}

Another way of deriving a similar result is to consider an approach related to the derivation of Bernoulli's equation. Take the scalar product of the momentum equation,  
\begin{equation}
\bar \rho \pd{\tilde v_i}{t} + \bar \rho \tilde v_j \pd{\tilde v_i}{x_j} = - \pd {\bar p}{x_i} 
- \pd{}{x_j} \langle \rho v_i^\prime v_j^\prime\rangle\textrm{,}
\end{equation}
with $\tilde v_i$,
\begin{equation}
\bar \rho \pd {}{t} \left( \frac{\tilde v^2}{2} \right) + \bar \rho \tilde v_j \pd{}{x_j} \frac{\tilde v^2}{2} = 
- \tilde v_i \pd {\bar p}{x_i} - \tilde v_i  \pd{}{x_j} \langle \rho v_i^\prime v_j^\prime \rangle \textrm{.} \label{eq:turbjetvmult}
\end{equation} 
Equation (\ref{eq:turbjetvmult}) describes the increase of the quantity $\tilde v^2 / 2$ under the action of the pressure gradient, gravitational force and turbulent diffusion. 
The gravitational term and the turbulent term $- \tilde v_i  (\partial / \partial x_j) \langle \rho v_i^\prime v_j^\prime\rangle $ reduce $\tilde v^2 $ so that the most optimistic acceleration is described by
\begin{align}
\bar \rho \frac{\dd}{\dd t} \left( \frac {\tilde v^2}{2} \right) & =  - \tilde v_i \frac{\partial\bar p}{\partial x_i} \\
 \Rightarrow v_2^2 - v_1^2 & \approx -2 \int_{z_1}^{z_2} \frac{1}{\bar \rho} \frac{\partial\bar p}{\partial z} dz\textrm{.} \label{speed}
\end{align}

The standard analysis of Bernoulli's equation assumes an equation of state for $p(\rho)$. In view of the complications of turbulent flow in this case, the relation between $\bar p$ and $\bar \rho$ would require a very detailed model. However, as with the momentum budget approach, all of the terms in equation~(\ref{speed}) can be estimated from the data, and the integration of the right-hand side can be performed numerically. The end result is the same as for the analysis based on the momentum budget. The pressure gradient fails by approximately two orders of magnitude to produce the increase in jet velocity observed.

\section{Acceleration of a Protostellar Jet by Embedded Magnetic Fields}\label{app:Baccel}

Consider the full expression for the energy flux density $\boldsymbol{F}_\textrm{E}$ carried by the jet,
\begin{equation}\label{eq:FE}
\boldsymbol{F}_\textrm{E} = \left(\frac{1}{2}v^2+h+\phi\right)\rho\boldsymbol{v} + \underbrace{\frac{B^2v}{4\pi}\left( \hat{\boldsymbol{v}}-\hat{\boldsymbol{v}}\cdot\hat{\boldsymbol{B}}\hat{\boldsymbol{B}} \right)}_\textrm{Poynting flux}\textrm{,}
\end{equation} 
where $\rho$ is the jet density, $\phi$ is the gravitational potential, $\boldsymbol{B}$ is the magnetic field, and hats denote unit vectors. Assuming a constant value of the jet energy flux across the jet cross-sectional area, $A_\textrm{jet}$, the \emph{total} jet power, $L_\textrm{jet}$, is then given by
\begin{align}
L_\textrm{jet}& = \boldsymbol{F}_\textrm{E}\cdot\hat{\boldsymbol{v}}A_\textrm{jet} \nonumber\\
              & = \left[\left(\frac{1}{2}v^2+h+\phi\right) + \frac{B^2}{4\pi\rho}\left(1-(\hat{\boldsymbol{v}}\cdot\hat{\boldsymbol{B}})^2\right)\right]\rho vA_\textrm{jet}\textrm{.}
\end{align}
If one assumes, as a first approximation, that both the total jet power and the jet mass flux, $\dot{M}=\rho vA_\textrm{jet}$, are constant\footnote{Strictly speaking, the total jet power will not be constant, as some energy must be radiated away as observable emission. However, this would affect the enthalpy term of equation (\ref{eq:FE}), which is typically negligible. Whilst this statement about enthalpy may not be true for post-shock regions in jet knots, it should be a good approximation for the non-shocked portion of the jet, which is the region observed to be accelerating.}, then one can form the equivalent of the Bernoulli equation for a hydromagnetic jet:
\begin{equation}\label{eq:magBern}
\left(\frac{1}{2}v^2+h+\phi\right) + \frac{B^2}{4\pi\rho}\left(1-(\hat{\boldsymbol{v}}\cdot\hat{\boldsymbol{B}})^2\right) = \frac{L_\textrm{jet}}{\dot{M}} = \textrm{const.}
\end{equation}

We consider three extreme cases of equation (\ref{eq:magBern}). First, if the magnetic field is parallel to the jet velocity, then the Poynting flux term disappears, and equation (\ref{eq:magBern}) collapses back to the purely hydrodynamic Bernoulli equation, equation (\ref{eq:Bern}), which has already been argued to be incapable of driving coupled acceleration-expansion in this scenario. Second, if the magnetic field is perpendicular to the jet velocity, then $\hat{\boldsymbol{v}}\cdot\hat{\boldsymbol{B}}=0$, and from the flux-freezing theorem, we deduce that the magnetic field of a self-similar jet evolves approximately as $B\propto 1/vR$, where $R$ is the jet radius. We next choose a reference point in the flow, and denote the values of magnetic field, density, velocity and radius at that point with a subscript zero. The jet magnetic field and density will then evolve thus:
\begin{align}
B    &= B_0\left(\frac{v}{v_0}\right)^{-1}\left(\frac{R}{R_0}\right)^{-1}\textrm{, and} \\
\rho &= \rho_0\left(\frac{v}{v_0}\right)^{-1}\left(\frac{R}{R_0}\right)^{-2}\textrm{.}
\end{align}
This leads to the expression
\begin{equation}
\frac{B^2}{4\pi \rho}=\frac{B_0^2}{4\pi\rho_0}\left(\frac{v}{v_0}\right)^{-1}\textrm{,}
\end{equation}
which has no $R$ dependence. Jet acceleration occurring in this regime would not show an increase in jet radius with jet velocity. Such an increase in radius is observed in the DG Tau jet (\S\ref{sec:D-bluepower}). Therefore, coupled acceleration-expansion cannot occur in this magnetic field configuration.

The third limiting case is that of a completely tangled magnetic field. Such a field behaves like a $\gamma=4/3$ gas, where $\gamma$ is the polytropic index of the gas, such that
\begin{equation}\label{eq:Bandrho}
\frac{B^2}{8\pi}\propto\rho^{4/3}\Rightarrow B^2 = B_0^2\left(\frac{\rho}{\rho_0}\right)^{4/3}\textrm{}
\end{equation}
\citep{KB04}. The Poynting flux term in equation (\ref{eq:magBern}) may be evaluated by assuming the velocity is in the outflow-axis direction only, and then averaging over solid angle, such that
\begin{equation}
\langle 1-(\hat{\boldsymbol{v}}\cdot\hat{\boldsymbol{B}})^2\rangle=\frac{2}{3}\textrm{.}
\end{equation}
Substituting the above into equation (\ref{eq:magBern}) yields the following equation relating quantities at a reference point, denoted by a subscript zero, to some other point along the outflow:
\begin{align}
\left(\frac{v}{v_0}\right)^2 &+ \frac{2(h-h_0)}{v_0^2}+\frac{2(\phi-\phi_0)}{v_0^2}-1 \nonumber\\
                             &= \frac{B_0^2}{3\pi\rho_0 v_0^2}\left[ 1-\left(\frac{v}{v_0}\right)^{-1/3}\left(\frac{R}{R_0}\right)^{-2/3}\right]\textrm{.} \label{eq:expaccel}
\end{align}
The enthalpy and gravitational potential terms of equation (\ref{eq:expaccel}) are generally unimportant in protostellar outflows at large distances from the central star.

\end{document}